\documentclass[sigconf,screen]{acmart}

\setcopyright{none}
\copyrightyear{2024}
\acmYear{2024}

\acmConference[]{ACM}{Survey}{PrivEx}

\usepackage{booktabs} %
\usepackage{mathtools}
\makeatletter \newif\if@restonecol \makeatother  
\usepackage[linesnumbered, ruled, vlined]{algorithm2e}
\usepackage{algpseudocode}

\usepackage[bottom]{footmisc}

\usepackage{epstopdf}
\usepackage{graphics}
\DeclareGraphicsExtensions{.eps,.gz,.eps,.pdf,.png,.jpg}

\graphicspath{{./}{./figure/}{fig/}}

\usepackage{xurl}
\usepackage{hyperref}

\newcommand\rurl[1]{%
  \href{https://#1}{\nolinkurl{#1}}%
}

\usepackage{makecell}
\usepackage{listings}
\usepackage{graphicx}
\usepackage{amsmath}
\usepackage{amsfonts}
\usepackage{amsthm}
\usepackage{dsfont}
\usepackage{wrapfig}
\usepackage{enumerate}
\usepackage{paralist}
\usepackage{multirow}
\usepackage{verbatim}
\usepackage{lipsum}
\usepackage{mathrsfs}
\usepackage{threeparttable}
\usepackage{tabularx}
\usepackage{subcaption}
\usepackage{adjustbox}

\theoremstyle{definition}

\theoremstyle{plain}
\newtheorem{example}{Example}

\newcommand{\edit}[1]{#1}
\newcommand{\sstitle}[1]{\smallskip\noindent\textbf{#1.\/}}

\def\Snospace~{\S{}}

\newcommand{\ie}{i.\,e.,\ }

\makeatletter
\newcommand{\removelatexerror}{\let\@latex@error\@gobble}
\makeatother

\hypersetup{
  pdftoolbar=true,        %
  pdfmenubar=true,        %
  pdffitwindow=false,     %
  pdfstartview={FitH},    %
  pdftitle={Complex Event Processing under Constrained Resources},    %
  pdfauthor={},     %
  pdfsubject={},   %
  pdfcreator={},   %
  pdfproducer={}, %
  pdfkeywords={}, %
  pdfnewwindow=true,      %
  colorlinks=true,       %
  linkcolor=Brown,          %
  citecolor=OliveGreen,        %
  filecolor=magenta,      %
  urlcolor=NavyBlue           %
}

 \newcommand\Mark[1]{\textsuperscript#1}

\begin{document}

\title[A Survey of Privacy-Preserving Model Explanations]{A Survey of Privacy-Preserving Model Explanations: \\ Privacy Risks, Attacks, and Countermeasures 
}

\author{
Thanh Tam Nguyen\Mark{1}, %
Thanh Trung Huynh\Mark{2},
Zhao Ren\Mark{3},
Thanh Toan Nguyen\Mark{1},
Phi Le Nguyen\Mark{4},
Hongzhi Yin\Mark{5}, %
Quoc Viet Hung Nguyen\Mark{1}%
}

\affiliation{%
  \institution{
  \Mark{1}Griffith University,
  \Mark{2}\'{E}cole Polytechnique F\'{e}d\'{e}rale de Lausanne,
  \Mark{3}University of Bremen,
  \Mark{4}Hanoi University of Science and Technology,
  \Mark{5}The University of Queensland
  }
  \country{}
}

\renewcommand{\shortauthors}{Thanh Tam Nguyen, et al.}

\begin{abstract}

As the adoption of explainable AI (XAI) continues to expand, the urgency to address its privacy implications intensifies. Despite a growing corpus of research in AI privacy and explainability, there is little attention on privacy-preserving model explanations. This article presents the first thorough survey about privacy attacks on model explanations and their countermeasures. Our contribution to this field comprises a thorough analysis of research papers with a connected taxonomy that facilitates the categorisation of privacy attacks and countermeasures based on the targeted explanations. This work also includes an initial investigation into the causes of privacy leaks. Finally, we discuss unresolved issues and prospective research directions uncovered in our analysis. This survey aims to be a valuable resource for the research community and offers clear insights for those new to this domain. To support ongoing research, we have established an online resource repository, which will be continuously updated with new and relevant findings. Interested readers are encouraged to access our repository at \url{https://github.com/tamlhp/awesome-privex}.

\end{abstract} 


\maketitle

\section{Introduction}
\label{sec:intro}

In recent years, the push for automated model explanations has gained significant momentum, with key guidelines like the GDPR highlighting their importance~\cite{goodman2017european}, and tech giants such as Google, Microsoft, and IBM pioneering this initiative by integrating explanation toolkits into their machine learning solutions~\cite{chang2021privacy}. This movement towards transparency encompasses a variety of explanation types, from global and local explanations that offer broad overviews and specific decision rationales, respectively, to feature importance analyses that pinpoint the impact of individual data inputs~\cite{ancona2018towards}. Techniques like SHAP and LIME provide nuanced insights into feature contributions~\cite{ribeiro2016should,lundberg2017unified}, while counterfactual explanations explore how changes in input could lead to different outcomes~\cite{guidotti2022counterfactual}. Additionally, interactive visualization tools are becoming increasingly popular, making the interpretation of complex models more accessible to users~\cite{bodria2023benchmarking,guidotti2018survey,gilpin2018explaining}.

However, this pursuit of transparency is not without its risks, especially privacy. The very act of providing explanations involves the disclosure of information that, while intended to illuminate, also carries the risk of inadvertently revealing sensitive details embedded in the models' training data. 
The balance between transparency and privacy becomes even more precarious when considering the granularity of explanations. Detailed explanations, although more informative, might offer direct inferences about individual data points used in training, thereby increasing the risk of privacy breaches. This paradox underscores a significant challenge within the field, as highlighted by recent research~\cite{goethals2023privacy,chang2021privacy,ferry2023sok}, which delve into the privacy implications of model explanations.

The degree to which model explanations reveal specifics about users' data is not fully understood. The unintended disclosure of sensitive details, such as a person's location, health records, or identity, through these explanations could pose serious concerns if such information were to be deciphered by a malicious entity~\cite{sokol2019counterfactual}. On the flip side, if private data is used without the rightful owner's permission, the same techniques aimed at exposing information could also detect unauthorized data utilization, thus potentially safeguarding user privacy~\cite{luo2022feature}. Furthermore, there is a growing interest not just in the attacks themselves but in understanding the underlying causes of privacy violations and what makes a model explanation susceptible to privacy-related attacks~\cite{naretto2022evaluating}. The leakage of information via model explanations can be attributed to a range of factors. Some of these factors are intrinsic to how explanations are crafted and the methodologies behind them, while others relate to the data's sensitivity and the granularity of the information the explanations provide~\cite{artelt2021evaluating}.

Given the paramount importance of protecting data privacy while simultaneously enhancing the transparency of machine learning (ML) models across domains, both the academic community and industry stakeholders are keenly focused on the privacy aspects of model explanations. To our knowledge, this article represents the inaugural comprehensive review of privacy-preserving mechanisms within model explanations. Through this work, we present an initial investigation that encapsulates both privacy breaches and their countermeasures in the context of model explanations, alongside explainable ML methodologies that inherently prioritize privacy. Furthermore, we develop taxonomies grounded in diverse criteria to serve as a reference for related research fields.

\begin{figure}[!h]
	\centering
	\includegraphics[width=0.5\linewidth]{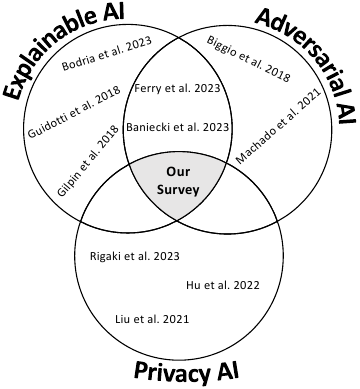}
\caption{This work vs. existing surveys. Explainable AI involves explanation and interpretable methods (e.g. \cite{bodria2023benchmarking,guidotti2018survey,gilpin2018explaining}). Adversarial AI includes adversarial attacks on ML models (e.g. \cite{machado2021adversarial,biggio2018wild}). Privacy AI involves privacy issues in ML (e.g. \cite{rigaki2023survey,hu2022membership,liu2021machine}). Others~\cite{ferry2023sok,baniecki2024adversarial} discuss exploits on model explanations. Our survey offers the first complete picture on privacy attacks, leaks, and defenses in explainable AI.}
	\label{fig:literature}
\end{figure}

\subsection{Comparisons with existing surveys}

Many surveys have summarised different privacy issues on ML models~\cite{biggio2018wild,papernot2017practical,machado2021adversarial,liu2021machine}, while others reviewed explanation methods for ML models~\cite{gilpin2018explaining,bodria2023benchmarking,adadi2018peeking}, but not both. For example, Rigaki et al.~\cite{rigaki2023survey} presented a thorough analysis of over 45 publications on privacy attacks in machine learning, spanning the last seven years. 
Hu et al.~\cite{hu2022membership} surveyed a special type of privacy attacks, called membership inference.
On the other hand, others~\cite{guidotti2018survey,adadi2018peeking,dovsilovic2018explainable} offered a comprehensive classification of model explanations to enhance interpretability and guided the selection of suitable methods for specific ML models and desired explanations.

Some existing surveys summarised adversarial attacks but presented partial coverage of privacy attacks on model explanations with basic introductions and limited discussions of the methods. Ferry et al.~\cite{ferry2023sok} examined the interplay between interpretability, fairness, and privacy, which are critical for responsible AI, particularly in high-stakes decision-making like college admissions and credit scoring. Baniecki et al.~\cite{baniecki2024adversarial} surveyed adversarial attacks on model explanations and fairness metrics, offered a unified taxonomy for clarity across related research areas, and discussed defensive strategies against such attacks.
However, these papers are either too high-level or too specialised in non-privacy attacks.

Our survey presents an in-depth examination of privacy attacks on model explanations, diverging from previous work by its comprehensive nature. Rather than addressing the full spectrum of adversarial attacks, our study is specifically tailored to privacy attacks. This focus is due to the recent surge in these attacks and their significant potential to compromise the right to explanation~\cite{goodman2017european} and the right to privacy~\cite{banisar2011right}. The threat posed by such privacy attacks could, in essence, challenge the very existence and usefulness of model explanations. Unlike the existing reviews that selected a very limited number of publications related to privacy attacks on model explanations (e.g. only two references are included in~\cite{baniecki2024adversarial}), we conduct a comprehensive search and include more than 50 related works in this survey. We delve into the underlying principles, theoretical frameworks, methodologies, and taxonomies, while also mapping out potential trajectories for future research. Especially, our work encompasses the emerging field of privacy-preserving explanations  (PrivEx), highlighting model explanations that inherently protect user privacy~\cite{vo2023feature,mochaourab2021robust,harder2020interpretable}.

\subsection{Paper collection methodology}

Finding relevant research on this subject proved to be complex due to its incorporation of various topics such as data privacy, privacy attacks, explanations of models, explainable AI (XAI), and the development of privacy-preserving explanations. To navigate this breadth of concepts, we employed diverse keyword combinations about ``privacy'', ``explanation'', and specific attack types including ``membership inference'', ``data reconstruction'', ``attribute inference'', ``model extraction'', ``model stealing'', ``property inference'', and ``model inversion''. Our initial search utilised platforms like Google Scholar, Semantic Scholar, and Scite.ai -- an AI-enhanced search tool -- to assemble a preliminary collection of studies. This selection was expanded through backward searches, analysing the references of initially chosen papers, and forward searches, identifying papers that cited the initial ones. Additionally, we manually verified the relevance and focus of these articles across various sources due to discrepancies, such as some studies addressing privacy in the context of safeguarding against manipulation attacks instead of privacy intrusions. Ultimately, this process culminated in nearly 50 pivotal research papers on the topic.

\subsection{Contributions of the article}

\begin{figure*}[!h]
	\centering
	\includegraphics[width=0.55\linewidth]{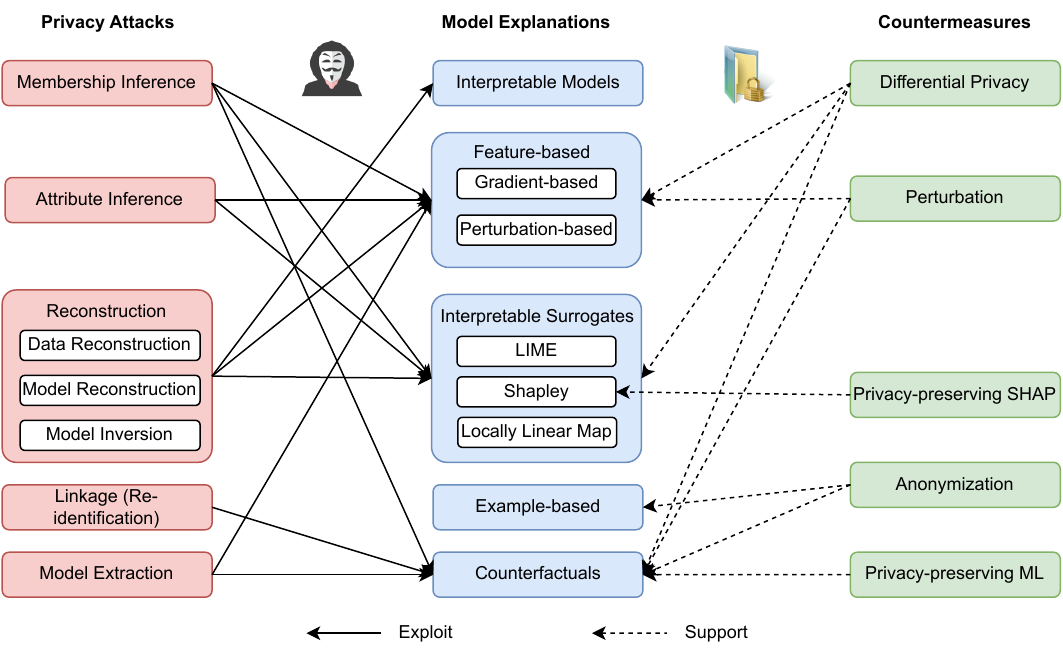}
\caption{Our taxonomy of privacy attacks and countermeasures on model explanations. ``Exploit'' arrows indicate existing works about privacy attacks on targeted explanations. ``Support'' arrows indicate existing works about privacy countermeasures for corresponding explanations. Some countermeasures (e.g. Privacy-Preserving ML) target privacy attacks directly and their arrows are omitted for brevity sake.}
	\label{fig:taxonomy}
\end{figure*}

The main contributions of this article are:

\begin{itemize}
	\item \textbf{Comprehensive Review:} To the best of our knowledge, this study represents the inaugural effort to thoroughly examine privacy-preserving model explanations. 
We have collated and summarised a substantial body of literature, including papers published or in pre-print up to March 2024. 

	\item \textbf{Connected Taxonomies:} 
We have organised all existing literature on PrivEx according to various criteria, including the types of explanations targeted and the methodologies employed in attacks and defences. \autoref{fig:taxonomy} showcases the taxonomy we have developed to structure these works.

	\item \textbf{Causal Analysis:} Recent research has begun to investigate conditions that could lead to privacy leaks through model explanations, indicating that some explanation mechanisms inherently possess vulnerabilities. To this end, we dedicate a section to discuss the probable causes of these leaks.

	\item \textbf{Challenges and Future Directions:} Designing privacy-preserving explanations for machine learning models is an emerging field of research. From the surveyed literature, we highlight unresolved issues and suggest several potential research directions into both the offensive and defensive aspects of privacy in model explanations. 
	
	\item \textbf{Datasets and Metrics:} In support of empirical research in PrivEx, we compile a comprehensive overview of datasets and evaluation metrics previously utilised in the field. 

	\item \textbf{Online Updating Resource:} To facilitate research in privacy-preserving model explanations, we have established an open-source repository\footnote{\url{https://github.com/tamlhp/awesome-privex}}, which aggregates a collection of pertinent studies, including links to papers and available code. 
\end{itemize}

\subsection{Organisation of the article}

The rest of the article is organised as follows. \autoref{sec:explanation} revisits model explanations, acting as foundations for privacy attacks. \autoref{sec:attack} presents the taxonomy of privacy attacks on model explanations and provides in-depth descriptions, including threat model and attack scenarios. \autoref{sec:risk} discusses the causes of privacy leaks in model explanations. \autoref{sec:defense} explores countermeasures and a new class of privacy-preserving model explanations by design. \autoref{sec:resource} provides the pinpoints to existing resources including source code, datasets, and evaluation metrics. Finally, \autoref{sec:future} contains a discussion on ongoing and upcoming research directions and \autoref{sec:conclusion} concludes the survey.

\section{Model Explanations}
\label{sec:explanation}

Model explanations serve to clarify the decisions a model renders concerning a specific querying sample denoted by $x$ represented as an n-dimensional feature vector ($x \in \mathbb{R}^n$). The explanation function $\phi$ ingests the dataset $D$, along with its labels -- either the ground truth labels $\ell: D \to [C]$ or those inferred by a trained model $f$ -- and the query $x \in \mathbb{R}^n$. Such methods for explanation may require access to supplementary data~\cite{chang2021privacy}, including the ability to query the model actively, a predefined notion of the data distribution, or familiarity with the class of the model~\cite{shokri2021privacy}. 

\autoref{tb:notation} summarises important notations in this paper.

\begin{table}[!h]
\centering
\caption{Summary of Important Notations.}
\label{tb:notation}
\footnotesize
\begin{adjustbox}{max width=1.0\linewidth}
\centering
\begin{tabular}{ll}
\toprule
 \textbf{Notation} & \textbf{Description} \\
  \midrule
$f: X \rightarrow Y$ & A machine learning model \\
$f_t$ & Target model of a privacy attack \\
$f_a$ & Adversarial model by a privacy attack \\
$D$ & Training data \\
$\phi (x) = \phi(D,f,x)$ & Explanation on the input data $x$ \\
$\phi^{GRAD}$(x) & Gradient-based explanation on input $x$ \\
$\phi^{INTG}$(x) & Integrated gradient-based explanation on input $x$ \\
$\phi^{SMOOTH}$(x) & Perturbation-based explanation on input $x$ \\
$\phi^{LIME}$(x) & LIME explanation on input $x$ \\
$\phi^{SHAP}$(x) & Shapley explanation on input $x$ \\
$\phi^{LLM}$(x) & Locally linear map-based explanation on input $x$ \\
$\phi^{CF}$(x) & Counterfactual explanation on input $x$ \\
$cf(x)$ & Counterfactual explanations/instances of the input data $x$ \\
$MI_{Distance}(x)$ & Distance-based membership inference attack on $x$ \\
$ \nabla_x f(x) $ & Gradient of the model $f$ on $x$ \\
$\hat{f}(.)$ & Surrogate model produced by model extraction attack \\
$\epsilon$-DP & Different privacy with $\epsilon$ degree or privacy budget \\
\bottomrule
\end{tabular}
\end{adjustbox}
\end{table}

\subsection{Feature-based Explanations}

The explanation function \( \phi(D, f, x; \cdot) \) is predicated on identifying influential attributes (with the \( \cdot \) symbol representing any potential additional inputs), and the explanation for the query $x$ is frequently referred to simply as $\phi(x)$~\cite{chang2021privacy}.
The value at the $i$-th index of a feature-based explanation, $\phi_i(x)$, quantifies the extent of influence the $i$-th feature exerts on the label ascribed to $x$. 
Ancona et al.~\cite{ancona2018towards} have curated a comprehensive exposition of these attribution-focused explanation modalities, also termed attribution methods or numerical influential measures~\cite{shokri2020exploiting}.

\begin{figure}[!h]
    \centering
    \includegraphics[width=.7\linewidth]{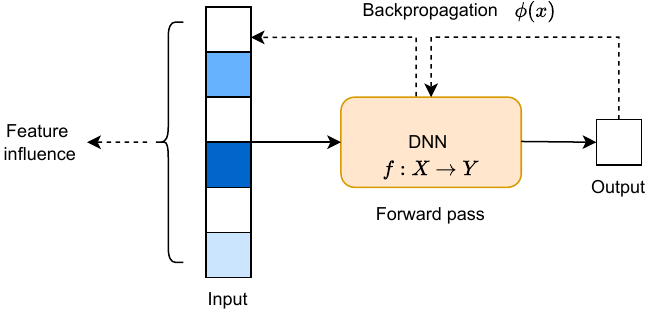}
    \caption{Feature-based explanations via backpropagation.}
    \label{fig:feature-explanation}
\end{figure}

\sstitle{Backpropagation-based (aka gradient-based)}
This type of explanation explains the decisions of neural network models through the lens of back propagation~\cite{shokri2021privacy} (see \autoref{fig:feature-explanation}). It allows for the allocation of the model's predictive reasoning back to the individual input features~\cite{simonyan2013deep,bach2015pixel,shrikumar2017learning,sliwinski2019axiomatic,smilkov2017smoothgrad,sundararajan2017axiomatic}.

\begin{itemize}
\item \emph{(Vanilla) Gradients:}
Simonyan et al.~\cite{simonyan2013deep} introduces gradient-based explanations, originally for image classification models, to emphasises important image pixels that affect the predictive outcomes. 
The explanation vector is defined as $\phi^{GRAD}(x) = \nabla_x f(x)$
or \( \phi_i({x}) = \frac{\partial f}{\partial x_i}({x}) \) for each feature $i$. A high partial differential value indicates that a pixel significantly affects the prediction, and analysing the map these values (so-called gradient map) can explain a model's decision-making~\cite{miura2021megex}.
Shrikumar et al.~\cite{shrikumar2017learning} suggest enhancing numerical explanations by using the input feature value multiplied by the gradient, \( \phi_i({x}) = x_i \times \frac{\partial f}{\partial x_i}({x}) \).

\item \emph{Integrated Gradients:}
Sundararajan et al.~\cite{sundararajan2017axiomatic} advocate for an alternative to standard gradient computation by averaging gradients along a straight path from a baseline input \( x^{BL} \) (often \( x^{BL} = \vec{0} \)) to the actual input. This method follows critical axioms like sensitivity and completeness. Sensitivity ensures that if there's a prediction change due to \( x_i \) not equaling \( x_{BL,i} \), then \( \phi_i({x}) \) should not be zero. Completeness dictates that the sum of all attributions equals the change in prediction from the baseline to the input.
\begin{equation}
\phi^{INTG}({x}_i) = (x_i - x_{BL,i}) \cdot \int_{\alpha=0}^{1} \frac{\partial c({x}^{\alpha})}{\partial x^{\alpha}_i} \bigg|_{{x}^{\alpha}={x} + \alpha({x}-x^{BL})}. 
\end{equation}

\item \emph{Guided Backpropagation:}
Designed for networks with ReLU activations (others as well), Guided Backpropagation~\cite{springenberg2014striving} modifies the gradient to only reflect paths with positive weights and activations, thereby considering only the positive evidence for a specific prediction. 

\item \emph{Layer-wise Relevance Propagation (LRP):}
proposed by Klauschen et al.~\cite{bach2015pixel} to assign relevance from the output layer back to the input features. The relevance in each layer is proportionally distributed according to the contribution from neurons in the previous layer. The final attributions for the input are referred to as \( \phi^{LRP}({x}) \).
\end{itemize}

\sstitle{Perturbation-based}
Perturbation-based explanations involve querying a model that needs to be explained with a series of altered inputs~\cite{shokri2021privacy}. 
SmoothGrad~\cite{smilkov2017smoothgrad} is a popular perturbation-based explanation method that produces several samples by injecting Gaussian noise into the input data and then computes the mean of the gradients from these samples.%
Formally, for a certain \( k \) samples, the explanation function is defined as:
\begin{equation}
 \phi^{\text{SMOOTH}}({x}) = \frac{1}{k} \sum_{k} \nabla_f({x} + \mathcal{N}(0, \sigma)),
 \end{equation}
where \( \mathcal{N} \) represents the normal distribution and \( \sigma \) stands for a hyperparameter that controls the level of perturbation.

\subsection{Interpretable Surrogates}
This method explains a black-box ML model or complex deep neural networks by computing a surrogate model that is interpretable by design~\cite{shokri2021privacy,deng2019interpreting,guidotti2018survey} that can emulate the overall predictive patterns of the original model~\cite{naretto2022evaluating}.

\sstitle{LIME}
Local Interpretable Model-agnostic Explanations~\cite{ribeiro2016should} generate a local interpretative approximation of a given model through sampling on the optimisation problem:
\begin{equation}
\phi^{\text{LIME}} (\bar{x}) = \arg\min_{g \in G} \mathcal{L}(g, f, \pi_{{x}}) + \Omega(g),
\end{equation}
where \( G \) is a collection of interpretable functions employed for explanatory purposes, \( \mathcal{L} \) quantifies how well \( g \) approximates \( f \) in the neighbourhood \( \pi_{{x}} \) of \( {x} \), and \( \Omega \) imposes a regularisation on \( g \) to avoid overfitting. Usually, $G$ involves one or multiple linear models and $\Omega$ is a Ridge regularisation~\cite{shokri2021privacy}.
The loss function is typically computed as the expected squared difference between the outputs of \( f \) and \( g \) weighted by the probability distribution \( \pi_{X} \)~\cite{slack2020fooling}:
\begin{equation}
L(f, g, \pi_{X}) = \sum_{x' \in X'} [f(x') - g(x')]^2 \pi_{X}(x')
\end{equation}
where $X'$ is the neighbourhood of $x$.

\sstitle{SHAP (local)}
The main distinction between LIME and SHAP is in the selection of the functions \( \Omega \) and \( \pi_x \). LIME takes a heuristic approach: \( \Omega(g) \) represents the count of non-zero weights within the linear model, while \( \pi_x(x') \) utilises either cosine or l2 distance~\cite{slack2020fooling}. 
SHAP values provide a way to quantify the contribution of each feature in a model prediction~\cite{jetchev2023xorshap,datta2016algorithmic,lundberg2017unified,vstrumbelj2014explaining,maleki2013bounding}. Specifically, for a given model \( f \) and a data point \( x = [x_1, \ldots, x_M] \), the SHAP value for feature \( i \) is calculated as a weighted average of differences between the model prediction with and without feature \( i \):
\begin{equation}
\phi^{SHAP}_i(x) = \sum_{S \subseteq \{1, \ldots, M\} \setminus \{i\}} \frac{1}{M} \frac{f_{S \cup \{i\}}(x) - f_S(x)}{{M-1 \choose |S|}}
\end{equation}
where \( |S| \) is the size of the subset \( S \) and \( M \) is the total number of features. 
For instance, let $x^0 = [x^0_i]_{i=1}^M$ be a reference sample of $M$ features. Suppose $M=4$, $x=[5,2,7,3]$, $x^0=[0,0,0,0]$, and we want to compute the marginal contribution $s_i$ of feature $i=1$ to the feature set $S=\{2,3\}$. Then $s_i = \frac{1}{4} \frac{f(x_{[1,2,3]}) - f(x_{[2,3]})}{3} = \frac{f([5,2,7,0]) - f([0,2,7,0])}{12}$.

\sstitle{Global Shapley Values}
The above Shapley values are local because the explanations are based on a singular reference sample $x^0$ and a single input sample $x$~\cite{slack2020fooling}. 
Begley et al.~\cite{begley2020explainability} proposes a Global Shapley Value by averaging local Shapley values over both foreground and background distributions, as given by:
\begin{equation}
\Phi^{SHAP}_i(f, F, B) = \mathbb{E} [\phi_i(f, x, x^0)]
\end{equation}
\edit{for each feature index \( i = 1, 2, \ldots, M \).
In other words, to conduct a global analysis of model behavior, it is necessary to consider predictions at multiple inputs \( x \sim \mathcal{F} \) from a distribution \( \mathcal{F} \) called the foreground. Since the choice of baseline \( x^0 \) is ambiguous, baselines \( x^0 \sim \mathcal{B} \) are sampled from a distribution \( \mathcal{B} \) called the background.
}

\sstitle{Locally Linear Maps}
Harder et al.~\cite{harder2020interpretable} introduces Locally Linear Maps (LLM), a method aimed at providing both local and global explanations for models, which is more expressive than standard linear models and offers an efficient way to manage the number of parameters for a good privacy-accuracy trade-off. 
\begin{equation}
 \phi^{LLM}_k(x) = \sum_{m=1}^{M} \sigma(x)^k_m g^k_m(x), \text{ where } g^k_m(x) = w^k_m \cdot x + b^k_m, 
 \end{equation}
and the weighting coefficients are computed via softmax:
\begin{equation}
\sigma^k_m(x) = \frac{\exp [\beta \cdot g^k_m(x)]}{\sum_{m=1}^{M} \exp [\beta \cdot g^k_m(x)]}. 
\end{equation}
The method optimizes a cross-entropy loss \( \mathcal{L}(W, \mathcal{D}) \) for the parameters of LLM collectively denoted by \( W \), with the predictive class label \( y_{n,k}(W) \) defined through a softmax function applied to the output of \( \phi_k(x_n) \).

\subsection{Example-based Explanations}

Example-based explanation (aka case-based interpretability or record-based explanation~\cite{shokri2020exploiting}) uses comparable examples to create transparent explanations for machine learning decisions, offering an accessible way to understand model predictions by contrasting similar cases from the model's database or generated data~\cite{montenegro2022privacy}. 
Case-based interpretability techniques can create a range of explanatory examples, including:

\begin{itemize}
	\item \emph{Similar examples:} are the closest matches from the training data with corresponding predictions to the case being analyzed, identified through a defined measure of similarity.
\item \emph{Typical examples:} representing the epitome of a particular prediction, frequently utilized in models that focus on prototype learning.
\item \emph{Counterfactual examples:} are similar examples but with differing predictions, highlighting the minimal changes needed for a different outcome. We dedicate a separate discussion on counterfactuals in the next subsection.
\item \emph{Semi-factual examples:} are similar to the original case with the same prediction but positioned near the decision boundary, demonstrating the robustness of the prediction against variations typical of a different classification.
\item \emph{Influential examples}: are key data points within a training set that have a significant impact on a model's prediction for a given query instance~\cite{koh2017understanding}. 
For explanatory purposes, we can provide the top \(k\) influential points~\cite{shokri2020exploiting}. 
\end{itemize}

These explanations can be sourced from existing datasets (i.e. $\phi(D,f,x;.) \in D$)~\cite{koh2017understanding} or crafted based on the original data~\cite{kenny2021explaining,lipton2018mythos}.

\sstitle{Intrinsic methods for traditional ML}
Case-based explanations in machine learning are derived from either distance-based or prototype-based interpretable methods. Distance-based methods utilize a measure of proximity to retrieve the most similar data points as explanations, while prototype-based methods classify and explain instances based on representative prototypes of clustered data. The K-Nearest Neighbors (KNN) algorithm exemplifies the former, offering explanations as similar or counterfactual examples based on label correspondence. The Bayesian Case Model (BCM) is a prototype-based method that explains decisions through typical examples representative of data clusters~\cite{kim2014bayesian}. Both methods aim to make model decisions understandable by referencing specific, characteristic data points or clusters~\cite{montenegro2022privacy}.

\sstitle{Posthoc methods for traditional ML}
Post hoc interpretability techniques leverage traditional machine learning models as metrics for finding similar examples, with decision trees and rule-based models used to determine similarity between data samples~\cite{montenegro2022privacy}. 
Counterfactual examples, on the other hand, come from nodes with differing outcomes. Moreover, models like Explanation Oriented Retrieval (EOR), built on the K-Nearest Neighbors (KNN) algorithm, reorder neighbors to highlight those with the highest explanatory utility, thus providing semi-factual examples that maintain the same classification but are closer to the decision boundary~\cite{nugent2009gaining}.

\sstitle{Intrinsic methods for deep learning}
In deep learning, intrinsic interpretability can be provided by prototype-based or distance-based methods~\cite{montenegro2022privacy}. For instance, the Explainable Deep Neural Network (xDNN)~\cite{angelov2020towards} and Deep Machine Reasoning (DMR)~\cite{angelov2020reasoning} define prototypes as dense data points and classify observations based on the closest prototype. 
The Prototype Classifier method learns representative prototypes from training data, using an autoencoder for feature extraction and classification based on latent representations~\cite{li2018deep}.%
The Prototypical Part Network (ProtoPNet) represents image parts in clusters in a latent space, which are used to predict and explain classifications~\cite{chen2019looks}. Additionally, the Deep k-Nearest Neighbors (DkNN) calculates neighbors at each model layer to ensure consistent predictions, offering explanations based on similar examples across the model's entirety~\cite{papernot2018deep}.

\sstitle{Posthoc methods for deep learning}
Post hoc interpretability methods in deep learning either utilise interpretable surrogate models to extract explanations from a primary model or directly analyse a ``black box'' model to identify anCSUx retrieve the most similar data instances for explanation purposes \cite{montenegro2022privacy}. Concept Whitening, for example, organises the latent space of a classification network around predefined concepts, enabling the measurement of distance between instances for similar example retrieval \cite{chen2020concept}. Interpretability guided Content-based Image Retrieval (IG-CBIR) enhances image retrieval by using saliency maps to focus on relevant image regions \cite{silva2020interpretability}. Unsupervised clustering and the KNN algorithm within the Twin Systems framework are other surrogate models that categorise or find similar examples based on feature extraction techniques like perturbation and sensitivity analysis \cite{kim2024does,kenny2019twin}.

\subsection{Counterfactual Explanations}

\begin{figure}[!h]
    \centering
    \includegraphics[width=0.7\linewidth]{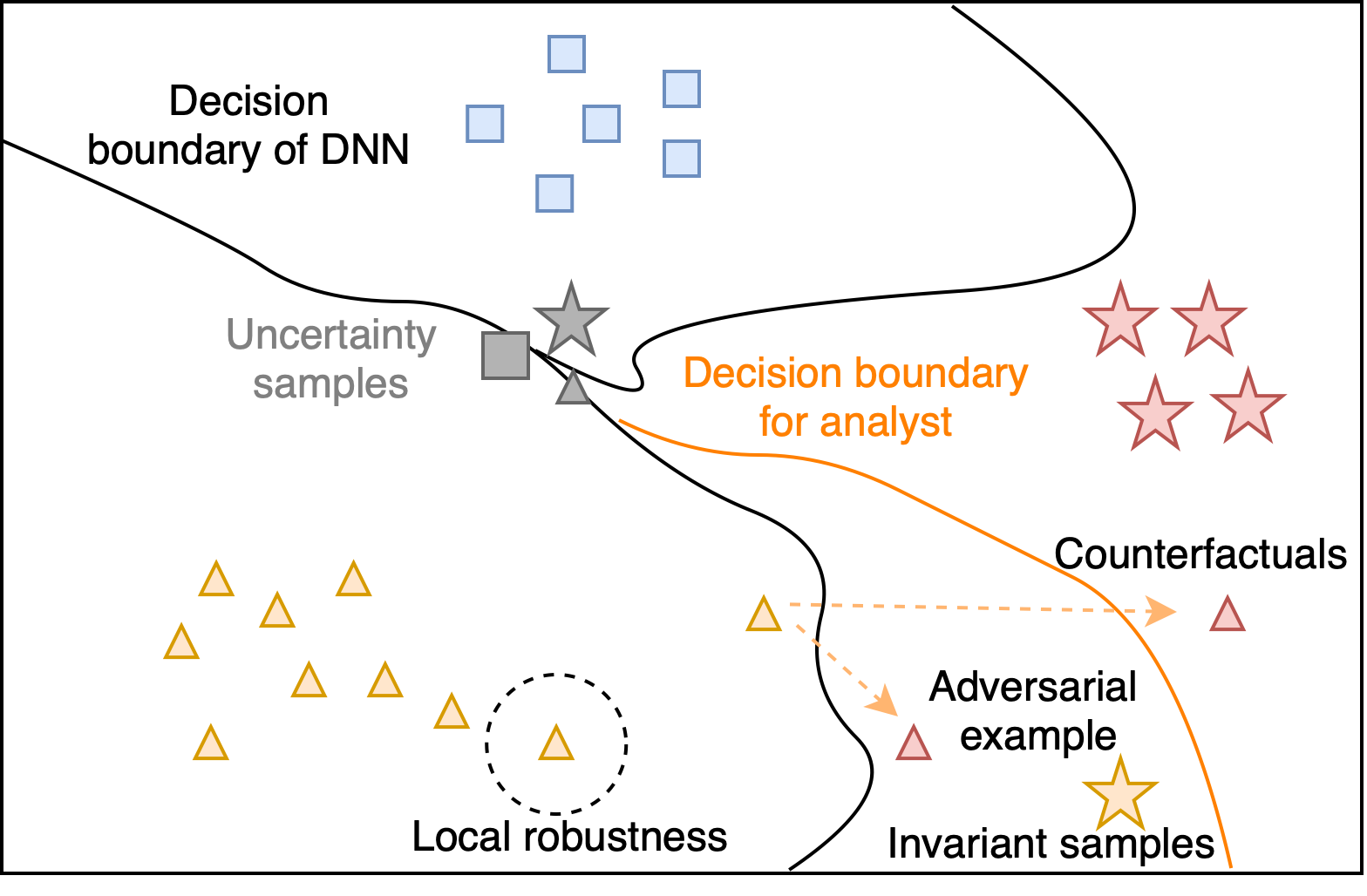}
    \caption{Decision boundaries between human analyst and a learnt model.}
    \label{fig:decisionboundary}
\end{figure}

Counterfactual explanations (aka algorithmic recourse) provide insights into how slight changes to input features could lead to different model outcomes, aiding in tasks like model debugging and ensuring regulatory compliance~\cite{goethals2023privacy,kuppa2021adversarial}. The study in~\cite{kuppa2021adversarial} gives an illustration of counterfactual and other four sample categories (\ie adversarial examples, local robustness~\cite{zhang2024does}, invariant samples, and uncertainty samples) through the boundaries between human analyst and a learnt model (see \autoref{fig:decisionboundary}).
The application of counterfactual explanations varies with the model's complexity and includes considerations such as model transparency, type compatibility, and adherence to constraints like feasibility and causality~\cite{wachter2017counterfactual,dodge2019explaining,binns2018s}. 
The concept overlaps with other areas of research such as algorithmic recourse, inverse classification, and contrastive explanations~\cite{karimi2021algorithmic,ustun2019actionable,laugel2017inverse,dhurandhar2018explanations}.

\sstitle{Single counterfactual}
Formally, counterfactual explanation is the process of finding changes \( \delta \) to an instance \( x \) that reverse a negative predictive outcome from a model \( f_\theta(x) = 0 \) to a positive one \( f_{\theta}(x + \delta) = 1 \), where $\theta$ are model parameters. The problem involves identifying a counterfactual \( x' = x + \delta \) where the predictive model outputs a positive outcome and doing so with minimal cost \( c(x, x') \), which is easily implementable, often using \( \ell_1 \) or \( \ell_2 \) distance as cost functions. 
The optimization problem is defined as:
\begin{equation}
\label{eq:counterfactual}
\phi^{CF}(x) = \text{arg min}_{x'\in A^P} L(f_{\theta}(x'), 1) + \lambda \cdot c(x, x')
\end{equation}
where $A^P$ is the set of plausible or actionable counterfactuals and $L(.,.)$ is a differential loss such as binary cross entropy~\cite{pawelczyk2023privacy}.

\begin{example}
	Possible counterfactual explanations derived from the FICO explainable machine learning challenge dataset~\cite{sokol2019counterfactual}:
	\begin{itemize}
		\item The model prediction for creditworthiness is negative. If the number of satisfactory trade lines had been 10 or fewer, rather than the actual 20, the prediction would have been positive.
		\item The model prediction for creditworthiness is negative. If there had been no trade lines that were ever 60 days overdue and marked as derogatory in the public record, rather than the actual count of 2, the prediction would have shifted to positive.
	\end{itemize}
\end{example}

\sstitle{Diverse counterfactuals}
Recent works study the generation of multiple alternative counterfactuals per input, offering a spectrum of  potential changes rather than just one nearest option~\cite{mothilal2020explaining}. This approach empowers users by offering them various ways they could potentially modify their data to achieve a preferred result~\cite{thang2015evaluation,nguyen2015tag,zhao2021eires}.

Kuppa et al.~\cite{kuppa2021adversarial} notes that methods for creating counterfactual explanations (CF) bear resemblance to those for generating adversarial examples (AE) in the way they both employ gradient-based optimization and surrogate models to find CF/AE for a given model. Some privacy attacks on adversarial examples can be used on counterfactual explanations~\cite{kuppa2021adversarial}.

\section{Privacy Attacks}
\label{sec:attack}

According to a classification system mentioned in~\cite{biggio2018wild,baniecki2024adversarial}, explainable AI systems can fall prey to three main categories of attacks: (i) integrity attacks, such as evasion and backdoor poisoning, leading to incorrect categorisation of certain data points~\cite{severi2021explanation,kuppa2020black,liu2022explanation,nguyen2023xrand}; 
(ii) availability attacks, characterised by poisoning efforts aimed at inflating the error rate in classification tasks~\cite{abdukhamidov2023hardening}; and (iii) privacy and confidentiality attacks, aimed at extracting sensitive information from user data and the models themselves. Although all forms of interference in machine learning can be considered adversarial, ``adversarial attacks" specifically denote those targeting the security aspect, particularly through malicious samples~\cite{garcia2018explainable,slack2020fooling,aivodji2022fooling,zhang2020interpretable}. 

This work is primarily concerned with breaches of privacy and confidentiality, including membership inference attacks, linkage attacks, reconstruction attacks, attribute/feature inference attacks, and model extraction attacks.
The rationale behind including model extraction attacks is their frequent association with privacy violations in related literature~\cite{rigaki2023survey}, and the notion that hijacking a model's functions could also infringe on privacy. Veale et al.~\cite{veale2018algorithms} contends that privacy violations like membership inference attacks elevate the likelihood of machine learning models being deemed personal data under the European Union's General Data Protection Regulation (GDPR), as they could make individuals identifiable.

\subsection{Membership Inference Attacks (MIA)}
MIA aim to detect if data is part of a model's training set~\cite{shokri2019privacy,shokri2021privacy}. Before model explanations, popular attacks are loss thresholding and likelihood ratio attack (LRT)~\cite{pawelczyk2023privacy}. 
Loss thresholding identifies if a data point was in the training set by checking the model's error rate against a threshold, requiring access to labels and model details~\cite{yeom2018privacy,sablayrolles2019white}. LRT, in contrast, uses shadow models to compare confidence levels of data being in or out of the training set, calculating a likelihood ratio to predict membership without needing direct model access~\cite{carlini2022membership}. 
Pawelczyk et al.~\cite{pawelczyk2023privacy} designs a recourse-based attack (using counterfactual explanation) without access to the true labels and knowledge of the correct loss functions.

\begin{figure}[!h]
    \centering
    \includegraphics[width=\linewidth]{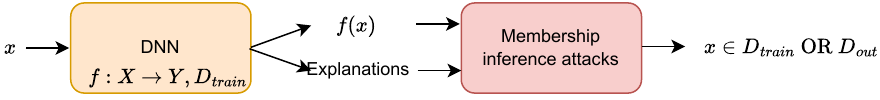}
    \caption{Membership inference attacks.}
    \label{fig:attack-membership}
\end{figure}

\sstitle{Threat model}
The adversary is able to submit $x$ to the black-box model~\cite{liu2022membership,li2022auditing,carlini2022membership,ye2022enhanced} to receive the prediction $f(x)$ and any corresponding explanations, despite not having direct access to the model's internals~\cite{quan2022amplification} (see~\autoref{fig:attack-membership}). 
However, they are assumed to know the model's architecture and possess an auxiliary dataset similar to the model's training data, reflected in much of the current research on the topic~\cite{liu2024please}.

\begin{itemize}
\item \emph{Threat model on gradient-based explanations:}
Most threat models are based on threshold-based attacks~\cite{shokri2021privacy}. There are two key scenarios for this: the optimal threshold scenario, where the threshold is deduced from known data point memberships to gauge the maximum privacy risk; and the reference/shadow model scenario, which is more practical and assumes the attacker has some labeled data from the same distribution as the target model, as well as knowledge of the model's architecture and hyperparameters in line with Kerckhoffs's principle~\cite{petitcolas2023kerckhoffs}. The attacker then trains a number of shadow models on this data to approximate the threshold, an approach that becomes more resource-intensive as the number of shadow models increases~\cite{shokri2021privacy}.

\item \emph{Threat model on interpretable surrogates:}
Naretto et al.~\cite{naretto2022evaluating} investigates how global explanation methods can potentially compromise the privacy. Specifically, the authors focus on TREPAN~\cite{craven1994using}, an algorithm that explains neural network decisions by creating a surrogate Decision Tree (DT) model. 

	\item \emph{Threat model on counterfactuals:}
Pawelczyk et al.~\cite{pawelczyk2023privacy} formulates a membership inference game for attacking counterfactual explanations. The game features two participants: a model owner ($\mathcal{O}$) and an opponent ($\mathcal{A}$). Their actions are as follows. $\mathcal{O}$ selects a dataset for training from a population \(D^N\), applying a training algorithm \(T\) with a loss function \(\ell\). 
Subsequently, $\mathcal{O}$ assigns a binary label \(f_\theta(z)\) to each datapoint \(z\) in \(D_t\). Let \(D_t^0\) be the segment of training data for which \(f_\theta(x) = 0\), and \(D_{\theta,0}\) represent the conditional distribution \(p(z) | f_\theta(z) = 0\). $\mathcal{O}$ tosses a coin, and based on the outcome, selects a sample \(x\) from either \(D_{\theta,0}\) or \(D_t^+\). Then, using the recourse algorithm $\phi$, $\mathcal{O}$ generates an alternate instance \(x'\) from \(\phi(f_\theta,x,D_t)\) and sends the pair \((x', x)\) to $\mathcal{A}$. In addition to the sample pair, $\mathcal{A}$ has the capability to make queries to \(D\). It is presumed that $\mathcal{A}$ is fully aware of $\mathcal{O}$'s implementation specifics, including the training algorithm \(T\) and the recourse algorithm \(\phi\). $\mathcal{A}$ concludes the game by providing a binary guess $G$ signifying if \(x\) belongs to \(D_t\) (MEMBER) or does not (\(x \notin D_t\), NON-MEMBER).
\end{itemize}

\sstitle{General attacks}
In the training set, data points are generally positioned away from the decision boundary, leading to lower loss scores that can be leveraged to detect membership in the training data~\cite{quan2022amplification,sablayrolles2019white,yeom2018privacy}. This principle is utilized in the OPT-var method~\cite{shokri2021privacy}, in which the variance in the explanation \( e=\phi(f, x) \) based on the logit score \( f(x) \) could signal whether a point was in the training set. However, Quan et al.~\cite{quan2022amplification} argues that logit scores alone may not fully represent the prediction confidence of the victim model because they do not take into account the scores of other classes. Instead, Quan et al.~\cite{quan2022amplification} suggests using the softmax function \( \sigma(f(x)) \), which reflects class interactions, to provide a more comprehensive membership indicator. 

\begin{figure}[!h]
    \centering
    \includegraphics[width=\linewidth]{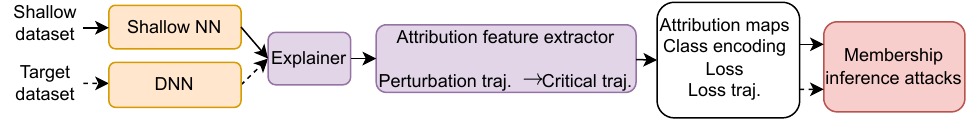}
    \caption{Model-based membership inference attacks proposed in~\cite{liu2024please}.}
    \label{fig:attack-membership-modelbased}
\end{figure}

Liu et al.~\cite{liu2024please} proposes a model-based attack that involves four main stages: training a shadow model, extracting attribution features, training an attack model, and inferring membership (see~\autoref{fig:attack-membership-modelbased}). The adversary starts by training a shadow model using an auxiliary dataset that is similar to the training data of the target model. Then, attribution maps are generated for a given sample, and perturbations are applied based on these maps to observe changes in predictions. 
Next, the adversary trains an attack model, typically a Multi-Layer Perceptron (MLP), using the attribution features combined with other data such as loss values and one-hot encoded class information to construct features indicative of membership.

\begin{itemize}
	
\item \emph{Attacks on gradient-based explanations:}
Shokri et al.~\cite{shokri2021privacy} uses a threshold-based attack that infers membership based on the model's confidence or its explanation output. A data point is classified as a member if the variance of the confidence scores $Var(f_\theta(x))$ or the variance of the explanation $Var(\phi(x))$ is below or equal to a certain threshold $\tau$.
Attacks using explanation variance exploit the model's certainty: when a model is sure about a prediction, explanation variance is low. However, near the decision boundary, even small changes can increase explanation variance. Models with certain activation functions like tanh, sigmoid, or softmax have steeper gradients, affecting how training data points are positioned relative to these boundaries~\cite{shokri2021privacy}.

\item \emph{Attacks on interpretable surrogates:}
Naretto et al.~\cite{naretto2022evaluating} develops an attacking procedure to assess the potential privacy risks of an interpretable surrogate (global explainer) that attempts to replicate the behavior of a black-box model.
First, an MIA model, denoted as \( A_b \), is trained to determine whether a specific data record, \( x \), was included in the training dataset, \( D_{train}^b \), of the black-box model \( b \). This attack model leverages the black-box \( b \) itself to classify the training data for the attack, making it specifically aimed at \( b \). The attack training dataset \( D_{train}^a \) is the same as \( D_{Attack}^B \). Similarly, another MIA model, \( A_c \), is developed to target the global explainer \( c \), which serves as an interpretable stand-in for the black-box model \( b \). This model is trained using \( D_{train}^a \), but this time the labeling is done by \( c \), not \( b \).

\item \emph{Attacks on counterfactual explanations:}
The adversary has access to both the original instance \( x \) and a counterfactual instance \( x' \). 
Models often overfit to training points, resulting in lower losses for these points compared to those on the test set~\cite{shokri2021privacy}.
Pawelczyk et al.~\cite{pawelczyk2023privacy} designs a distance-based attack where if the loss is below a certain threshold \( \tau \), the point is considered a MEMBER of the training set.
The counterfactual distance \( c(x, x') \) is effectively the distance to the model boundary, and even though algorithms that produce realistic recourses may not optimize for this distance, it can still be viewed as an approximation to the distance to the model boundary~\cite{karimi2021algorithmic,pawelczyk2020learning}.
The counterfactual distance-based attack is defined by \( MI_{Distance} (x) \) as follows:
\begin{equation}
MI_{Distance} (x) = 
\begin{cases} 
\text{Member} & \text{if } c(x, x') \geq \tau_D \\
\text{Non-member} & \text{if } c(x, x') < \tau_D
\end{cases}
\end{equation}
Another attack is using a Likelihood Ratio Test on top of the Counterfactual Distance (CFD)~\cite{pawelczyk2023privacy}. 
The process involves calculating a baseline statistic \( t_0 \) using \( c(x, x') \) from the recourse output. 
If the initial statistic \( t_0 \) surpasses the critical threshold \( z_{1-\alpha} \), which is the \( 1-\alpha \) quantile of the normal distribution \( Z \), the algorithm designates the data point as a `Non-member'; and `Member' otherwise. 
The key benefit is that it estimates the parameters \( \mu_{out}, \sigma_{out} \) only once for the non-membership scenario, reducing the computational load when assessing multiple data points \( x' \)~\cite{sablayrolles2019white}.

Huang et al.~\cite{huang2023accurate} proposes a CFD-based Likelihood Ratio Test (LRT) for linear classifiers built on the above Pawelczyk method~\cite{pawelczyk2023privacy}. 
But the attack is simplified and one-sided as it only estimates parameters for data outside the training set, thus reducing computational complexity. 

Kuppa et al.~\cite{kuppa2021adversarial} develops an attack that leverages an auxiliary dataset \( D_{aux} \) to train a shadow model \( A_{MemInf} \).
This is done by generating counterfactual examples \( x_{cfi} \) for input samples \( x_i \) and training a 1-nearest neighbor (1-NN) classifier to predict class membership based on proximity to these counterfactuals. If the prediction probability difference between the shadow model \( A_{MemInf} \) and the target model \( T \) is below a threshold \( t \), the sample is deemed part of the training set. This inference is made under the assumption that if both models predict similarly for a sample, it implies the sample was significant in its prediction. The method is advantageous as it requires no direct access to the training set and iteratively uses counterfactuals to extract new data.
\end{itemize}

\subsection{Linkage Attacks}

\begin{figure}[t]
    \centering
    \includegraphics[width=\linewidth]{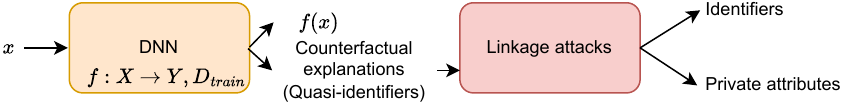}
    \caption{Linkage attacks.}
    \label{fig:attack-linkage}
\end{figure}

\sstitle{Threat model}
Goethals et al.~\cite{goethals2023privacy} introduces a privacy concern with counterfactual explanations when they are based on training instances. The data usually consist of identifiers (like name and social security number), quasi-identifiers (like age, zip code, gender), and private attributes. 
It has been shown that a significant portion of US citizens could be uniquely identified by combining their zip code, gender, and date of birth~\cite{sweeney2000simple}. 
The attack setup assumes the adversary has access to identifiers and quasi-identifiers. There are two re-identification scenarios discussed: one where a specific individual is targeted to uncover their private attributes, and another where the adversary aims to prove that re-identification is possible, regardless of who the individual is. Counterfactual explanations, which do not include identifiers but may contain unique combinations of quasi-identifiers, could be exploited by an attacker to infer private attributes in what is termed an ``explanation linkage attack'' or "re-identification attack"~\cite{goethals2023privacy} (see~\autoref{fig:attack-linkage}).

\sstitle{Attacks on counterfactual explanations}
Goethals et al.~\cite{goethals2023privacy} presents a scenario where Lisa is denied credit and requests a counterfactual explanation, which inadvertently reveals Fionas' private information because Fiona is the nearest unlike neighbor in the dataset. 
Native counterfactuals, which are real instances from the dataset, are more plausible but increase the risk of re-identification~\cite{brughmans2023nice}. Perturbation-based counterfactuals, which synthetically generate explanations, pose less privacy risk but can still be vulnerable to sophisticated attacks if the perturbations are minor~\cite{artelt2021evaluating,keane2020good,pawelczyk2020counterfactual}. 
Aivodji et al.~\cite{aivodji2020model} identifies that diverse counterfactual explanations can inadvertently expose decision boundaries more, risking the leak of sensitive data like health or financial information. Linkage attacks exploit this by matching anonymised records with external datasets, combining various attributes to re-identify individuals.

\subsection{Reconstruction Attacks}

\begin{figure}[!h]
    \centering
    \includegraphics[width=\linewidth]{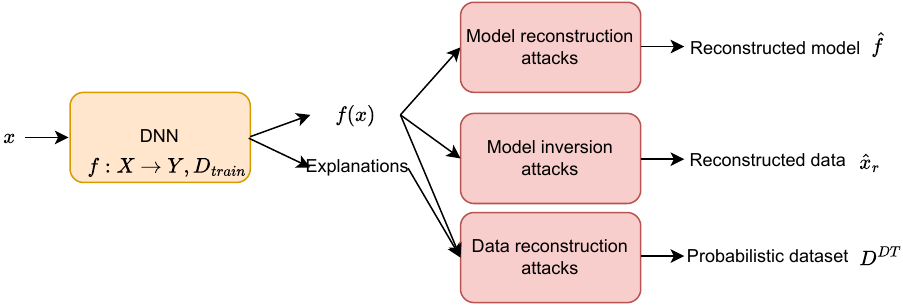}
    \caption{Reconstruction attacks.}
    \label{fig:attack-reconstruction}
\end{figure}

Based on model predictions and explanations, reconstruction attacks involve dataset reconstrcution attacks, model reconstruction attacks, and model inversion attacks (see~\autoref{fig:attack-reconstruction}).

\sstitle{Dataset reconstruction attacks}
It is important to preserve privacy in datasets due to several threats posed by inference attacks that seek to deduce sensitive information from model outputs~\cite{dwork2017exposed,rigaki2023survey}. Ferry et al.~\cite{ferry2023probabilistic,ferry2023addresing} reviews the evolution of reconstruction attacks from databases to machine learning, where adversaries attempt to recover training data. Techniques range from linear programming to exploiting data memorisation, even within frameworks meant to promote fairness~\cite{garfinkel2019understanding,song2017machine}. The goal of data reconstruction attacks is to make models trained for fairness inadvertently reveal sensitive attributes, including leveraging auxiliary datasets and queries to an auditor for enhancing attacks~\cite{carlini2019secret,salem2020updates}.

\begin{itemize}
	\item \emph{Threat model:} 
A machine learning model that is interpretable, like a decision tree, contains implicit information about its training dataset~\cite{ferry2023probabilistic}. This information can be formalized into a probabilistic dataset \( \mathcal{D} \) consisting of \( n \) examples, each with \( d \) attributes. Every attribute \( a_k \) has a domain \( V_k \) covering all possible attribute values. The knowledge about an attribute \( a_k \) for a given example \( x_i \) is represented by a probability distribution across all possible values for that attribute, using the random variable \( \mathcal{D}_{i,k} \). 
If a value \( \mathcal{D}_{i,k} \) within \( V_k \) has all the probability mass (i.e., \( P(\mathcal{D}_{i,k} = v_{i,k}) = 1 \)), it's deterministic. 
Conversely, a probabilistic dataset encompasses some uncertainty about attribute values.

\item \emph{Probabilistic Reconstruction Attacks:}
Earlier research~\cite{gambs2012reconstruction} proposes a method for constructing a probabilistic dataset \( \mathcal{D}^{DT} \) from the structure of a trained decision tree \( DT \). This probabilistic dataset reflects the decision tree's implicit knowledge about its training dataset \( \mathcal{D}^{Orig} \). The construction of this dataset is termed a probabilistic reconstruction attack, and by design, \( \mathcal{D}^{DT} \) is compatible with \( \mathcal{D}^{Orig} \), meaning the actual value \( v_{i,k}^{Orig} \) of any attribute \( a_k \) for any example \( x_i \) is always among the set of possible values in the probabilistic reconstruction (\( P(\mathcal{D}_{i,k}^{DT} = v_{i,k}^{Orig}) > 0 \)).

\item \emph{Attacks on Interpretable Models:}
Ferry et al.~\cite{ferry2023probabilistic} discusses the possibility of a probabilistic reconstruction attack on interpretable models.
In the general case, the success of the attack is calculated using the joint entropy of the dataset's cells, which can be simplified if the variables of the model are statistically independent. For interpretable models like decision trees and rule lists, this assumption allows further decomposition of the computation~\cite{ferry2023probabilistic}. 
\end{itemize}

\sstitle{Model reconstruction attacks}
Model reconstruction is the process of replicating a classifier \( \hat{f} \) when provided with membership and gradient queries to an oracle that, for any input \( x \), reveals both the classifier's output \( \hat{f}(x) \) and the gradient \( \nabla_x \hat{f}(x) \). Milli et al.~\cite{milli2019model} examines a specific scenario involving a one hidden-layer neural network function \( f : \mathbb{R}^d \rightarrow \mathbb{R} \) that uses ReLU activations, formulated as \( f(x) = \sum_{i=1}^{h} w_i \max(A_i^T x, 0) \). 

\begin{itemize}
	\item \emph{Threat model:}
	For a DNN with parameters \( A \in \mathbb{R}^{h \times d} \) and \( w \in \mathbb{R}^h \), where \( A_i \) represents the ith row of A, three assumptions are posited: (1) Each row \( A_1, ..., A_h \) is a unit vector; (2) No pair of rows \( A_i \) and \( A_j \) are collinear for \( i \neq j \), satisfying \( \langle A_i, A_j \rangle \leq 1 - c \) for some \( c > 0 \); (3) The rows \( A_1, ..., A_h \) are linearly independent. These assumptions are stated to be without loss of generality since they can be achieved by simple reparameterization of the network, such as scaling \( w \) or \( A \), or by reducing the hidden layer dimension.

\item \emph{General attacks:}
Under these assumptions, it is possible to learn the function with a sample complexity independent of the input dimension \( d \)~\cite{milli2019model}. Specifically, with a probability of \( 1 - \delta \), an algorithm can find a function \( \hat{f} \) such that \( \hat{f} = f \). If the algorithm cannot find such a function, it will report the failure. Regardless of the outcome, the algorithm requires only \( O\left(h \log \frac{h}{\delta}\right) \) queries to learn the function.

	\item \emph{Attacks on gradient-based explanations:}
The algorithm involves recovering a matrix \( Z \) and a sign vector \( s \)~\cite{milli2019model}. The matrix \( Z \) is composed of either \( w_i A_i \) or \( -w_i A_i \), with the signs encapsulated in \( s \). The function \( f \) can then be reconstructed from \( Z \) and \( s \), utilizing the recovered structure to make predictions. 
The approach relies on exploiting the gradient structure of \( f \) to identify the hyperplanes that partition the input space and uses binary search to recover the necessary components of \( Z \) and \( s \). 

\end{itemize}

\sstitle{Model inversion attacks}
Model inversion attacks aim to deduce original data from predictions, such as recreating a person's face based on their predicted emotional state~\cite{fredrikson2015model,yang2019neural,zhang2020secret}. 
Initially, model inversion attacks showed limited success~\cite{fredrikson2015model}, but advancements in deep learning, especially through the use of transposed Convolutional Neural Networks (CNNs), have significantly enhanced their effectiveness~\cite{dosovitskiy2016inverting,he2019model,yang2019neural}. Additional enhancements have been achieved by utilising auxiliary information, including access to the model's internal workings and feature embeddings, or understanding the joint probability distribution between features and labels~\cite{zhang2020secret,yeom2018privacy,he2019model}. Especially the increasing demand for model explanations is likely to make these attacks more common~\cite{zhao2021exploiting}.

\begin{itemize}
	\item \emph{Threat model:} We consider a machine learning model \(f_t\) that processes confidential data \(x\) from a set \(X_p\) (for instance, facial images). It employs these private inputs to generate a prediction \(\hat{y}_t\) (such as identifying emotions).
	An issue arises when an attacker gains access to the target prediction \(\hat{y}_t\) and the explanation $\phi_t$ (due to reasons like a data breach, interception during transmission, or sharing on social media). 
One scenario is to assume that the attacker only has the compromised data, an independent dataset \(x \in X_a\), and the ability to interact with the target model via black-box~\cite{zhao2021exploiting}. The attacker does not require additional privileged information, such as blurred versions of the images.
The objective of the attacker is to develop their own inversion model \(f_a\) to reconstruct the original image \(x\) from the model's outputs \(\hat{y}_t, \phi_t\)). Such a reconstruction would allow them to predict sensitive information from the reconstructed image \(\hat{x}_r\), including the possibility of re-identifying the individual from the facial emotion recognition system~\cite{hu2022protecting}.

\item \emph{Attack on a single gradient-based explanation:} 
To invert the target model $M_t$, a Transposed Convolutional Neural Network (TCNN)~\cite{dumoulin2016guide} is devised to reconstruct a two-dimensional image $x_r$ from the one-dimensional prediction vector $y_t$ provided by $M_t$. The TCNN minimises the mean squared error (MSE) loss to approximate the original image. This TCNN incorporates various input forms, such as saliency maps and 2D explanations~\cite{selvaraju2017grad,simonyan2013deep}, enhancing the reconstruction of $x_r$. Inputs can be processed by flattening the 2D explanations into a 1D vector and concatenating with the prediction vector, or by using a CNN to convert 2D patterns into a 1D feature embedding, following the approach used in CNN encoder-decoder networks and super-resolution techniques~\cite{ur2019end,zhang2020secret}. A U-Net architecture is employed to improve the reconstruction fidelity~\cite{zhang2018residual}.
A hybrid model that combines flattened explanations with the U-Net structure is introduced in~\cite{zhao2021exploiting}. 
The training objective for these models is defined by the image reconstruction loss function:
\begin{equation}
L_r = \sum_x (M^{a}_i (M_t (x)) - x)^2
\end{equation}
where $x$ represents the original image, $M_t(x) = y_t$ denotes the prediction from the target model, and $M^{a}_i (M_t(x)) = x_r$ is the reconstructed image output. Zhao et al.~\cite{zhao2021exploiting} conducts experiments on how different explanation methods, including gradients~\cite{simonyan2013deep}, CAM~\cite{zhou2016learning}, LRP~\cite{bach2015pixel}, and blurred versions of the input images, affect the inversion model's ability to capture information. 

\item \emph{Attack on multiple gradient-based explanations:}
While many explanations clarify the reasons a model predicts a certain class within a set $C$, it is equally crucial to elucidate why it did not predict a different class $c' \neq c$, offering contrastive insights~\cite{miller2019explanation}. To facilitate this, certain techniques like Grad-CAM can generate explanations that are specific to a class based on the user's query~\cite{selvaraju2017grad}. Nevertheless, this approach increases the risk to privacy as it provides additional information. Zhao et al.~\cite{zhao2021exploiting} makes use of these Alternative CAMs ($\Sigma$-CAM) by merging explanations across all classes in \( |C| \) into a three-dimensional tensor, and they train their inversion models on this tensor rather than on a two-dimensional matrix representing a single explanation. 

\item \emph{Attack on surrogate explanations:}
Interpretable surrogates could be harnessed for inversion attacks, even for models that do not provide target explanations. Zhao et al.~\cite{zhao2021exploiting} proposes an attack that predicts the target explanation and exploits that explanation to invert the original target data. Initially, an explainable surrogate target model $f_a$ is trained using the attacker's dataset to generate a surrogate explanation \( \widetilde{\phi} \). 
However, \( \widetilde{\phi_t} \) is only accessible during the training phase and not during prediction. Consequently, an explanation inversion model \( f_e \) is trained to reconstruct \( \widetilde{\phi_t} \) as \( \widehat{\phi_r} \) based on the target prediction \( \widehat{y_t} \). The proposed loss function for minimising the surrogate explanation error is:
\begin{equation}
L_\phi = \sum_x \left( f_e (f_t (x)) - \phi(f_t (x)) \right)^2
\end{equation}
where \( \phi(f) \) denotes the explanation of the model \( f \), \( {f_t (x)} = {y_t} \) represents the surrogate target prediction, \( \phi(f_t (x)) = \widetilde{\phi_t} \) is the surrogate explanation, and \( f_e (f_t (x)) = \widehat{\phi_r} \) is the reconstructed surrogate explanation. This reconstructed explanation is available at prediction time. Finally, \( \widehat{\phi_r} \) is fed into the image inversion model \( \phi_i \) to finalize the model inversion attack. 
Given that \( \widehat{\phi_r} \) is formatted similarly to \( \widetilde{\phi_t} \), any explanation methods can be applied.

\item \emph{Attacks on confidence scores:} 
Fredrikson et al.~\cite{fredrikson2015model} develops a model inversion attack
by using a maximum a posteriori (MAP) estimator to compute \( f(x_1, \ldots, x_d) \) for all possible values of the sensitive feature \( x_1 \), while exploiting confidence information from model predictions.
Fredrikson et al.~\cite{fredrikson2015model} addresses the challenge of inverting high-dimensional features like facial recognition, where the inversion task becomes an optimization problem solved by gradient descent. 
\end{itemize}

\begin{figure}[!h]
    \centering
    \includegraphics[width=\linewidth]{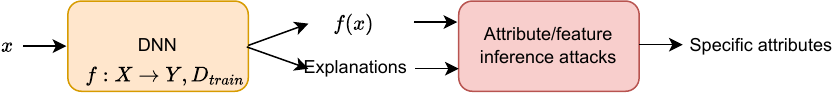}
    \caption{Attribute/feature inference attacks.}
    \label{fig:attack-attribute}
\end{figure}

\subsection{Attribute/Feature Inference Attacks}

Attribute inference attacks, aka feature inference attacks, are designed to deduce specific attributes, such as gender, from individual data records by using accessible data like model predictions or explanations~\cite{song2019overlearning,yeom2018privacy} (see~\autoref{fig:attack-attribute}). 
These types of attacks are distinct from property inference attacks, which seek to ascertain broader dataset characteristics, like the training data's gender ratio~\cite{ganju2018property,melis2019exploiting,zhang2021leakage}.

Duddu et al.~\cite{duddu2022inferring} investigates a scenario where a machine learning model, \( f_{target} \), is cloud-deployed within an MLaaS framework (e.g. Google Cloud, Microsoft Azure), capable of providing predictions and required explanations for any given input. 
Users can submit a private sample \( x = \{x_i\}^n_{i=1} \) to the service provider and receive a prediction vector \( \hat{y} = \{\hat{y}_i\}^c_{i=1} \), along with an explanation vector \( \phi = \{\phi_i\}^n_{i=1} \) that pertains to a specific class. Although the service provider has the capacity to return multiple explanation vectors corresponding to different classes~\cite{chen2018learning}, for practicality and without loss of generality, most works focuses on the use of one explanation vector for a specific class~\cite{luo2022feature}.

\sstitle{Threat models on feature-based explanations} 
Duddu et al.~\cite{duddu2022inferring} considers two threat models (TM). (1) \emph{TM1} (with \( s \) in \( D \)): Here, the sensitive feature \( s \) is included in both the training dataset \( D \) and the input. \( \mathcal{A}dv \) has access to the predictions \( f_{target}(x \cup s) \) and explanations \( \phi(x \cup s) \), but not the ability to pass inputs to the model. The adversary's goal is to train an attack model \( f_{adv} \) that maps the explanations \( \phi(x) \) to \( s \) on \( D_{aux} \), an auxiliary dataset known to \( \mathcal{A}dv \). 
(2) \emph{TM2} (without \( s \) in \( D \)): In this scenario, \( s \) is not included in the dataset \( D \) or the input \( x \). 
Unlike TM1, \( \mathcal{A}dv \) can pass inputs \( x \) to the model and has blackbox access to \( f_{target} \) and \( \phi(x) \), making this a more practical threat where \( s \) is censored for privacy. \( \mathcal{A}dv \)'s goal remains the same, to infer \( s \) by training \( f_{adv} \) on \( D_{aux} \).
For both models, the adversary has an additional auxiliary dataset \( D_{aux} \) that contains data records with non-sensitive and sensitive attributes along with their corresponding labels.

\sstitle{Threat models on Shapley values}
Unlike previous assumptions~\cite{salem2018ml,shokri2021privacy} that adversaries have an auxiliary dataset with a distribution similar to the target sample, Luo et al.~\cite{luo2022feature} explores two relaxed scenarios. The first adversary has access to an explanation vector, an auxiliary dataset, and a black-box prediction model, aiming to reconstruct the target sample. The second adversary operates under more practical constraints with only black-box access to the machine learning services and the explanation vector, without any background knowledge of the target sample.

\sstitle{Attacks on feature-based explanations}
Duddu et al.~\cite{duddu2022inferring} develops an attribute inference attack based on thresholding. The attack model, \( f_{adv} \), uses model explanations to infer sensitive attributes and chooses the threshold \( t^* \) that maximizes the F1-Score. This calibration step deviates from using the typical default threshold of 0.5 to increase the precision and recall of the attack, particularly when there is a moderate to large class imbalance of the sensitive attribute \( s \).
Duddu et al.~\cite{duddu2022inferring} also shows low Pearson correlation coefficients between the sensitive attribute \( s \) and other entities like \( y \), \( x \), and \( \phi(x) \) across different datasets and explanation methods, suggesting little to no direct correlation between the sensitive attribute and the model's predictions or explanations, challenging the notion that the attack is merely exploiting these correlations.

\sstitle{Attacks on Shapley values}
Luo et al.~\cite{luo2022feature} proposes an attack where an adversary, with access to a black-box model \( f \), attempts to infer private input features from Shapley value explanations. To simplify the computation of Shapley values, the adversary uses a reference sample \( x^0 \) and a linear transformation function \( h \). They aim to reduce mutual information between the input \( x_i \) and the Shapley value \( s_i \) to zero, meaning the adversary cannot gain any information about \( x_i \) from \( s_i \). Luo et al.~\cite{luo2022feature} assumes that the Shapley values follow a Gaussian distribution, and thus the probability \( P(s_i) \) is modelled as a Gaussian function.
To ensure that the mapping from the auxiliary input data \( X_{aux} \) to the Shapley values \( S_{aux} \) is bijective, Luo et al.~\cite{luo2022feature} presents a theorem requiring \( X_{aux} \) to be finite. The adversary can then use a hypothesis \( \psi \) to map Shapley values back to the auxiliary input data. To execute the attack, the adversary collects the Shapley values for all \( x_{aux} \in X_{aux} \), sends prediction queries to the MLaaS platform, and obtains explanations \( S_{aux} \). They then train a regression model on \( X_{aux} \) to learn the mapping \( \psi \) from Shapley values \( S_{aux} \) to \( X_{aux} \).

Another scenario is where an adversary lacks an auxiliary dataset to carry out a feature inference attack~\cite{luo2022feature}. Without knowledge of the target's data distribution, it becomes challenging to learn an attack model by observing Shapley values. 
To mitigate these challenges, the adversary can use the linear correlation between feature values and Shapley values for important features. By drawing samples independently and using a Generalized Additive Model (GAM) for approximation, the adversary can restore features from Shapley values. 
Luo et al.~\cite{luo2022feature} notes that while their attacks work well with Shapley values, other explanation methods like LIME and DeepLIFT may not be suitable due to their heuristic-based, unstable mappings between features and explanations.

\subsection{Model Extraction Attacks}

\begin{figure}[!h]
    \centering
    \includegraphics[width=\linewidth]{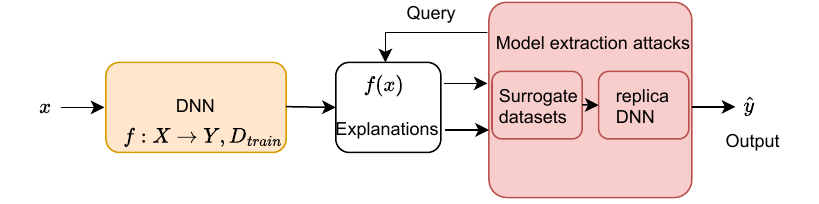}
    \caption{Model extraction attacks.}
    \label{fig:attack-model-extraction}
\end{figure}

There is an increasing concern of model extraction attacks in the context of Machine Learning as a Service (MLaaS)~\cite{tramer2016stealing}, where attackers steal ML models by using surrogate datasets to make queries through the MLaaS API, and then train replica models with the obtained predictions. The goal is to create a functionally equivalent version with identical predictions (see~\autoref{fig:attack-model-extraction}). The difference between a model extraction attack and a model reconstruction attack is that the former does not need to know the model architecture.

Research on model extraction attacks targeting explainable AI systems is emerging~\cite{mi2024towards}. Milli et al.~\cite{milli2019model} develops a method that leverages the discrepancy in gradient-based explanations between an original AI model and its clone, demonstrating enhanced attack efficiency. Additionally, Ulrich et al.~\cite{aivodji2020model} designs an attack utilising counterfactual explanations to train a cloned model with greater effectiveness. Miura et al.~\cite{miura2021megex} designs a data-free attack that does not require surrogate datasets in advance.

\sstitle{Threat models}
An adversary duplicates a trained model, referred to as the victim model \( f : X \to Y \), by utilising its predictions to create a similar clone model \( \hat{f} : X \to Y \). The adversary's goal is to replicate the victim model's accuracy using only the output predictions. On the one hand, typical model extraction attacks~\cite{milli2019model} involve the adversary collecting input data \( x \in X \), querying the victim model to obtain predictions, and using the pairs \( (x_i, f(x_i)) \) to compile a dataset for training the clone model. 
In some scenarios, an adversary requires query access to the victim model but does not necessarily need the training data's ground-truth labels~\cite{quan2022amplification}. The attack relies on knowing the architecture of the victim model but not its parameter values. The attacker aims to produce a model that performs identically on the same test dataset, although the adversary's extracted model may not have been trained on the same data or in the same manner as the victim model.

On the other hand, data-free model extraction attacks~\cite{miura2021megex} eliminates the need for input data collection, in which an adversary employs a generative DNN \( G : \mathbb{R}^r \to X \) to convert Gaussian distribution noise into synthetic input data. The adversary then uses this data to query the victim model and gather training pairs \( (x, f(x)) \), which are used to train the clone model to emulate the victim model \( f \). The generative model is designed to create data that, when predicted by the clone model, is different from the victim model's output, intending to maximize the clone model's loss function and improve parameter updates. Although the generative model \( G \) does not learn the actual distribution of the input data space \( X \), it is optimised to produce data that facilitates the clone model's training process.

In the case of counterfactual explanations, the explanation API provides for each data point \( x_i \), a corresponding counterfactual explanation \( c(x_i) \), accompanied by the predicted outcome \( \hat{y}_i \). When seeking a collection of diverse counterfactuals, the API will yield a collection \( C(x_i) \) comprising multiple counterfactual instances, rather than just a single example.

\sstitle{Attacks on gradient-based explanations}
In the data-free model extraction~\cite{miura2021megex}, an attacker crafts a surrogate model, denoted as \( \hat{f}: X \rightarrow Y \), alongside a generative model \( G: \mathbb{R}^r \rightarrow X \), responsible for creating synthetic data inputs. An iterative process is repeated between two steps. The first step generates \( N_G \) input samples and queries the target model to refine the generative model based on both predictions and explanations, utilising these explanations to compute the gradient \( \nabla_{\theta_G} \mathcal{L} \). The second routine produces \( N_C \) input samples for querying the target model and uses the resulting predictions to train the surrogate model. The process stops when the number of queries ($N_G + N_C$) aligns with the allocated query budget \( Q \). This strategy enables the attacker to leverage the gradient \( \nabla G(x) = \nabla_x f(x) \) for the training of this generative model.

Adversarial attacks for model extraction without data rely on alternately calculating the gradients of an objective function with the parameters of both a cloned model and a generative model. Training the clone requires calculating the gradient $\nabla_{\theta_f} \mathcal{L}$, achievable via back-propagation by the adversary. However, current methods do not provide the adversary with access to $\nabla_{\theta_G} \mathcal{L}$ for training the generative model. According to~\cite{miura2021megex}, it suffices to find $\nabla_x \mathcal{L}(x)$ as it leads to $\nabla_{\theta_G} \mathcal{L} = -\nabla_{\theta_G} G(z) \cdot \nabla_x \mathcal{L}(x)$
Unlike previous methods that only provided terms other than $\nabla_x f(x)$, the adversary now gains explanations through the standard Gradient $G(x) = \nabla_x f(x)$, enabling the computation of $\nabla_x \mathcal{L}(x)$ precisely. The adversary can employ almost any differentiable loss function for training the generative model.

Quan et al.~\cite{quan2022amplification} proposes another explanation-matching attack~\cite{milli2019model}, focusing on replicating both the predictions and explanations of the original, or victim, model. The adversary's model minimises two losses: the prediction loss (the difference in predictions between the two models) and the explanation matching loss (the difference in their explanations). The overall loss being minimised is a weighted combination of these two losses. Additionally, the method includes the use of LIME to ensure the interpretability of predictions matches that of the victim model.

\sstitle{Attacks on counterfactual explanations}
Kuppa et al.~\cite{kuppa2021adversarial} considers two main factors: (a) The auxiliary dataset \( D_{aux} \) should approximate the training set of \( f \). This can be challenging if \( D_{aux} \) does not naturally follow the training distribution, but counterfactual explanations can provide samples from various classes that may bridge this gap. An attacker can iteratively query and obtain diverse class samples to better reflect the training set distributions. (b) Knowing the architecture of \( f \) can significantly enhance the fidelity of the extracted model. However, in realistic scenarios, attackers often lack this information, complicating the attack.
To circumvent this obstacle, once data samples that mirror the training set are collected, knowledge distillation techniques are employed. This involves transferring insights from \( f \) to a surrogate model \( g \). The knowledge transfer is quantified using a distillation loss, given by \( L_{Distill}(f, g) = L_{KL}(P_f(x), P_g(x)) \), where \( L_{KL} \) represents the Kullback-Leibler divergence loss. In this setup, the attacker leverages publicly available data and queries \( f \), then applies the distillation loss to train \( g \), thereby extracting the functionality of \( f \).

Aivodji et al.~\cite{aivodji2020model} proposes a model extraction attack~\cite{jagielski2020high} by 
compiling an attack set and training a surrogate model on the collected data from counterfactual samples.
Counterfactual explanations typically change features with larger importance values to achieve the desired prediction, thus revealing the model's sensitive areas. However, this approach has limitations, such as the decision boundary shift issue caused by using distant queries from the decision boundary as training samples~\cite{aivodji2020model}. This leads to an unstable substitute model and requires more queries to resolve, thus increasing the attack cost. Wang et al.~\cite{wang2022dualcf} proposes a method called DualCF to mitigate this issue by using pairs of counterfactuals (CF) and their corresponding explanations (CCF) from the opposite class as training data. This helps to balance the substitute model's decision boundary and improve extraction efficiency.
DualCF for a Linear Model is also discussed, illustrating that for binary linear models, it's possible to extract a substitute model with 100\% agreement using CF and CCF pairs. 
While promising for linear models, extending this approach to nonlinear and complex models remains a challenge, and the effectiveness of DualCF in those scenarios is yet to be thoroughly evaluated~\cite{tramer2016stealing}.

\section{Causes of Privacy Leaks}
\label{sec:risk}

Research into the causes that lead to privacy leakage through model explanations has started to emerge in the past few years~\cite{naretto2022evaluating,shokri2021privacy,artelt2021evaluating,chang2021privacy,pawelczyk2023privacy,quan2022amplification}.  Certain types of explanations are prone to divulging data, often due to their inherent structure. For instance, case-based explanations, which utilise actual data points from the training set, can inadvertently reveal sensitive information~\cite{montenegro2022privacy,shokri2020exploiting}. Other explanations, such as surrogate models (e.g. SVM, linear classifiers) are relative easy to leak their parameters by querying enough input/output data pairs~\cite{naretto2022evaluating,quan2022amplification,ferry2023probabilistic}.

\subsection{Privacy Leaks in Counterfactual Explanations}

While counterfactual explanations aim to clarify AI decisions, they may inadvertently compromise privacy~\cite{sokol2019counterfactual}. These explanations can give adversaries clues to manipulate the system, as seen in instances where absence of a feature (like a savings account) leads to a better outcome than a suboptimal presence~\cite{sokol2019counterfactual}. 
They provide insights into decision boundaries, potentially revealing model specifics and training data, such as feature splits in logical models, training points in k-nearest neighbors, or support vectors in SVMs. 
Moreover, the existence of multiple and varying-length counterfactuals for a single data point could increase the ease of model theft, with longer, more complex counterfactuals potentially disclosing substantial model information with just one explanation.

Vo et al.~\cite{vo2023feature} outlines essential privacy concepts relevant to public datasets. Identifiers are personal attributes capable of uniquely distinguishing an individual, such as names or government-issued numbers. Quasi-identifiers, while not individually unique, can collectively re-identify individuals; a mix of gender, birthdate, and ZIP code, for instance, can pinpoint 87\% of American residents~\cite{sweeney2000simple}. Sensitive attributes cover confidential information like salaries or medical records that need safeguarding to prevent personal or emotional harm. 
To protect against re-identification risks, public datasets need to undergo anonymisation by removing direct identifiers, though vulnerability remains due to quasi-identifiers.

\begin{example}
	In the given scenario from the FICO explainable ML dataset~\cite{sokol2019counterfactual}, the outcome of the credit evaluation could have shifted from negative to positive if one of the following conditions were met: 
	\begin{itemize}
		\item \# installment trades is less than 3 instead of \textbf{3}
		\item \# revolving trades is less than 3 instead of \textbf{5}
		\item \# trades with 60 days overdue and marked as derogatory in public record is equal to 0 instead of \textbf{2}.
		\item \# loans within 1 year is less or equal to 2 instead of \textbf{5}.
	\end{itemize}
	Here, user privacy is violated as the exact values of the above sensitive attributes are revealed~\cite{sokol2019counterfactual}.
\end{example}

Diverse counterfactuals equip users with a range of actionable insights to potentially alter their outcomes favorably~\cite{mothilal2020explaining,nguyen2023feasible}. However, this also increases privacy risks as it may give away additional details that could be exploited for more potent attacks~\cite{aivodji2020model}.
Artelt et al.~\cite{artelt2021evaluating} identifies a key problem with counterfactual explanations: their instability to minor input variations can lead to significantly different outcomes for similar cases. Addressing this, the authors propose studying the robustness of counterfactual explanations and suggest using plausible rather than closest counterfactuals to enhance stability~\cite{artelt2020convex}.

\subsection{Causes of Membership Inference Attacks}
Membership inference attacks (MIAs) aim to predict whether a data point is in the training set or not~\cite{shokri2020exploiting}. The trade-off between explainability and privacy has been investigated and evaluated using membership inference attacks in \cite{naretto2022evaluating,shokri2021privacy,chang2021privacy,pawelczyk2023privacy}. 

\sstitle{Global explainers} 
Naretto et al.~\cite{naretto2022evaluating} demonstrates that interpretable tree-based global explainers can increase the risk of privacy leakage. To explain $f$, an interpretable global surrogate classifier $g$ is required to be trained to imitate the behavior of $f$, \ie $g(X)=f(X)$. 
To compare the privacy exposure risk caused by $f$ and $g$, two attack models are trained: one is learnt by querying $f$, and the other queries $g$. It was found that the global explainer is more vulnerable to the membership inference attack model than the classifier~\cite{naretto2022evaluating}, resulting in more privacy exposure.

\sstitle{Feature-based explanations} MIAs were also evaluated on feature-based explanations, including back-propagation and perturbation \cite{shokri2021privacy}. 
Backpropagation-based explanations were found to result in privacy leakage, which may be caused by high variances of explanations. A high variance of an explanation indicates that the point is close to the decision boundary and has an uncertain prediction, which is helpful for an adversary.  Compared to backpropagation-based explanations, perturbation-based explanations are more robust to membership inference attacks. This might be because the query points are not used to train the model~\cite{shokri2021privacy}.

\sstitle{Repeated interaction}
Kumari et al.~\cite{kumari2024towards} focus on repeated interactions. The author introduce attacks using explanation variance to infer data membership, modeled through a continuous-time stochastic signaling game. The study proves an optimal attack threshold exists, analyzes equilibrium conditions, and uses simulations to assess attack effectiveness in dynamic settings.

\sstitle{Fairness}
Apart from explanations, pursuing fairness during model training can also increase risks of privacy exposure~\cite{chang2021privacy}. When processing imbalanced data, fairness constraints require the model to memorize the training data in the smaller groups rather than learning a general pattern~\cite{chang2021privacy}. Such a way makes it easier for membership inference attacks to attack the model. Especially, when membership inference attacks are designed specifically for each group, they showed higher attack accuracy than that of a common membership inference attack for all groups~\cite{chang2021privacy}. Another study~\cite{shokri2020exploiting} also reports small groups in record-based explanations are more vulnerable to membership inference attacks than majority groups. 

\sstitle{Influence of Input Dimension}
Shokri et al.~\cite{shokri2021privacy} evaluates how the input dimension influences the privacy risks of gradient-based explanations. 
Their experiments revealed that 
as the number of features grows (between $10^3$ and $10^4$), a correlation between gradient norms and training membership appears, indicating vulnerability to membership inference attacks. 
However, this effect is moderated by the number of classes and is also dependent on model behavior, as overfitting can occur with too many features. While increasing the number of classes generally increases learning problem complexity, the actual impact on the correlation between gradient norms and membership depends on the specific range of features and that the interval and amount of correlation vary.

\sstitle{Influence of Overfitting}
Yeom et al.~\cite{yeom2018privacy} demonstrates that overfitting has a notable impact on the success of membership inference attacks. Shokri et al.~\cite{shokri2021privacy} conducts tests varying the number of training iterations to achieve different levels of accuracy, in order to assess the effects of overfitting. Consistent with prior research on loss-based attacks, they found that their threshold-based attacks, which leverage explanations, are more effective when targeting overfitted models.

\subsection{Causes of Reconstruction Attacks}
Reconstruction attacks target on reconstructing the partial or complete training data. 
Ferry et al.~\cite{ferry2023probabilistic} shows that post-hoc explanations can disproportionately impact individual privacy, exacerbating risks for minority groups. This trend towards reduced privacy for minorities is also reflected in interpretability, as identified by Shokri et al~\cite{shokri2021privacy,shokri2020exploiting,shokri2019privacy}. They discovered that the likelihood of discerning whether an individual's data was used in a model's training set from post-hoc explanations is higher for outliers and certain minority groups that the model finds difficult to generalize. This increased risk is attributed to these groups being more frequently included in the generated explanations. Consequently, tools designed for interpretability could inadvertently lead to greater information leakage about these already vulnerable groups.

Interpretable models enhance transparency but can inadvertently disclose information about their training data. Gambs et al.~\cite{gambs2012reconstruction} uses such data leakage to probabilistically reconstruct a decision tree's training set. The uncertainty within this reconstruction can be measured to determine how much information the model leaks.

Ferry et al.~\cite{ferry2023probabilistic,ferry2023addresing} examines how optimal and heuristic decision trees and rule lists reveal information about their training data. 
The study finds that optimal models tend to leak less information than greedily-built ones for a given level of accuracy. It also notes significant variance in how much information individual training examples contribute to the overall entropy reduction, with some examples inherently leaking more information based on their position within the model's structure.

\subsection{Causes of Property Inference Attacks}

Regularisation techniques like dropout and ensemble learning have been shown to prevent models from memorizing private inputs, potentially reducing the risk of information leakage~\cite{luo2021feature,melis2019exploiting,liu2022efficient}. Despite previous findings, Luo et al.~\cite{luo2022feature} reveals that incorporating dropout in neural networks at varying rates (0.2, 0.5, 0.8) actually enhances the accuracy of certain attacks. This counterintuitive result is attributed to dropout preventing overfitting by smoothing the decision boundaries, which inadvertently benefits the attack. Nevertheless, a very high dropout rate (0.8) does decrease the success rates of one attack due to underfitting and increased randomness in the model, which disrupts the linearity between inputs and outputs.

Case-based explanation methods, often used in sensitive fields like medical diagnosis, risk privacy breaches when they share detailed visual data with unauthorized viewers, such as medical students or family members~\cite{montenegro2022privacy}. 
To mitigate this, anonymisation techniques must be applied to the images before they are shared, ensuring that the identity of individuals is not disclosed while still preserving the explanatory power and realism of the images. 
The anonymisation process involves altering identity features in the latent vector to produce a privatized image, but there's no guarantee that other latent features don't inadvertently reveal identity, especially if facial embeddings capture significant identifiable information.

\subsection{Causes of Model Extraction Attacks}

Quan et al.~\cite{quan2022amplification} explores how model extraction attacks can benefit from explanation methods, leading to adversarial gains with fewer queries. A particular finding is that while certain explanation methods, such as Gradient, Integrated Gradient, and SmoothGrad, can be exploited to enhance attack efficiency, others like Guided Backprop and GradCam may result in poorer performance due to biases in gradient estimation.

While counterfactual explanations (CFs) do not reveal the entirety of a cloud model's workings, their impact on security and privacy has been underestimated~\cite{barocas2020hidden,kasirzadeh2021use,sokol2019counterfactual}. Some research argues that CFs only unveil a minimal amount of information, showing a limited set of dependencies for an individual instance which might seem insufficient for model extraction~\cite{hashemi2020permuteattack,wachter2017counterfactual}. However, accumulating enough data through multiple queries can significantly facilitate the extraction process~\cite{wang2022dualcf}. 
Aivodji et al.~\cite{aivodji2020model} pioneers the use of model extraction attacks on counterfactual explanations by treating these explanations near decision boundaries as supplementary training data.
Wang et al.~\cite{wang2022dualcf} also shows that adversaries can exploit CF explanations to extract a high-fidelity model by learning about the decision boundaries.

\subsection{Causes of Explanation Linkage Attacks}

Vo et al.~\cite{vo2023feature} reviews key concepts relevant to data privacy, specifically in the context of public datasets. Identifiers are attributes that can uniquely identify an individual, like names or government numbers. Quasi-identifiers, while not unique on their own, can combine to uniquely identify a person. Sensitive attributes are confidential data that, if disclosed, could harm an individual. 
Public datasets are at risk of explanation linkage attacks, aka re-identification attacks, even after anonymisation if quasi-identifiers are present~\cite{vo2023feature}. 
Their experiments acknowledge that k-anonymity lower the risks but it may still allow private information to be inferred through homogeneity and background knowledge attacks.

\section{Privacy-Preserving Explanations}
\label{sec:defense}

\subsection{Defences with Differential Privacy}

Differential privacy (DP) is a solid, mathematically based privacy standard that defines privacy loss using a quantifiable metric~\cite{liu2024matrix}. It does so through mechanisms that guarantee the aggregated data output will obscure the involvement of any individual record in the dataset, as established by Dwork et al.~\cite{dwork2014algorithmic}. Differential privacy is usually formalized as follows~\cite{huang2023accurate}. A randomized mechanism \( M \) with domain \( D \) and range \( R \) achieves \( \varepsilon \)-differential privacy (\( \varepsilon \)-DP) if, for all adjacent datasets \( d, d' \) differing by one row, and for any output set \( S \subseteq R \), the following inequality holds: 
\begin{equation}
 \text{Pr}[Q(d) \in S] \leq e^\varepsilon \cdot \text{Pr}[Q(d') \in S]. 
 \end{equation}
Here, \( \varepsilon \) is the privacy loss parameter, where smaller values correspond to stronger privacy.

\begin{figure}[!h]
    \centering
    \includegraphics[width=\linewidth]{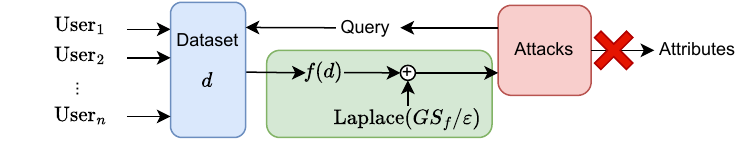}
    \caption{Differential Privacy.}
    \label{fig:defence-differential}
\end{figure}

The Laplace Mechanism of differential privacy is useful for queries on numerical data~\cite{huang2023accurate}. As shown in \autoref{fig:defence-differential}, the mechanism adds noise to the sensitive query's output according to the Laplace distribution. Specifically, for a sensitive query function \( Q(d) \), the \( \varepsilon \)-DP Laplace Mechanism \( Q_{Lap} \) is given by \( Q_{Lap}(d) = Q(d) + \text{Laplace}(GS_Q / \varepsilon) \), where \( \text{Laplace}(GS_Q / \varepsilon) \) represents a random variable from the Laplace distribution with a scale dependent on the global sensitivity \( GS_Q \) divided by \( \varepsilon \). Global sensitivity \( GS_Q \) is the maximum norm-1 difference of \( Q \) across all pairs of adjacent datasets \( d, d' \). Lastly, Dwork et al.~\cite{dwork2014algorithmic} have demonstrated a post-processing property of differential privacy:
If \( Q \) is \( \varepsilon \)-DP and \( G \) is any arbitrary deterministic mapping, then the composite function \( G \circ Q \) is also \( \varepsilon \)-DP~\cite{huang2023accurate}.

\subsubsection{Differentially Private Feature-based Explanations}
An explanation \( \phi(\cdot) \) is \( (\epsilon, \delta) \)-differentially private if the probability of any sequence of explanations does not change significantly with the addition or removal of a single data point in the training set~\cite{patel2022model}. For a sequence of queries \( \vec{z}_1, ..., \vec{z}_k \), and any two neighboring training sets \( \mathcal{D} \) and \( \mathcal{D}' \), and subsets \( S_1, ..., S_k \subseteq \mathbb{R}^n \), we have:
\begin{equation}
Pr[\phi^1 \in S_1, ..., \phi^k \in S_k] \leq e^\epsilon \cdot Pr[\phi'^1 \in S_1, ..., \phi'^k \in S_k] + \delta
\end{equation} 
where \( \phi^i = \phi(\vec{z}_i, f_{\mathcal{X}}(\vec{x})) \) and \( \phi'^i = \phi(\vec{z}_i, f_{\mathcal{D}'}(\vec{x})) \) for all \( i \). The privacy for the explanation dataset \( \mathcal{X} \) can follow a similar guarantee.
Despite these measures, post-hoc explanation algorithms, which are applied after the model has been trained, cannot fully prevent membership inference attacks, since they do not control the training process or parameters~\cite{patel2022model}. 

\sstitle{Single explanation algorithm}
Patel et al.~\cite{patel2022model} focuses on creating differentially private feature-based model explanations, where \( \phi(\vec{z}) \) is a vector in \( \mathbb{R}^n \) that quantifies the impact of each feature on the model's predicted label \( f_{\mathcal{D}}(\vec{z}) \). The aim is to find a local explanation function \( \phi \), centred at a point of interest \( \vec{z} \), that minimises the local empirical model error over an explanation dataset \( \mathcal{X} \). The local empirical loss of \( \phi \) over \( \mathcal{X} \) is given by:
\begin{equation}
\mathcal{L}(\phi, \vec{z}, f_{\mathcal{X}}) = \frac{1}{|\mathcal{X}|} \sum_{\vec{x} \in \mathcal{X}} \alpha(\|\vec{x} - \vec{z}\|)(\vec{x} - \vec{z})^T(\vec{x} - \vec{z}) - f_{\mathcal{X}}(\vec{x})^2, 
\end{equation}
where \( \alpha \) is a weight function that decreases with distance from \( \vec{z} \).
The optimal model explanation is the one that minimises this loss:
\begin{equation}
\phi^*(\vec{z}, f_{\mathcal{X}}) = \arg\min_{\phi \in \mathcal{C}} \mathcal{L}(\phi, \vec{z}, f_{\mathcal{X}}).
\end{equation} 

To ensure differential privacy, Patel et al.~\cite{patel2022model} introduces a Differentially Private Gradient Descent (DPGD) algorithm, which utilises the Gaussian mechanism to protect the explanation dataset \( \mathcal{X} \). The privacy of the explanation dataset is protected by computing a private version of the gradient descent updates. The DPGD-Explain procedure iteratively updates \( \phi \) using the gradient of the loss function perturbed by Gaussian noise, aiming to find the minimum of \( \phi \) within a certain bound:
\begin{equation}
 \phi^{(t+1)} \leftarrow \arg\min_{\phi \in \mathcal{C}_2,1} \| \phi - \zeta^{(t)} \|,
 \end{equation}
where \( \zeta^{(t)} \) is the perturbed gradient at iteration \( t \).
Patel et al.~\cite{patel2022model} provides conditions for bounded sensitivity for the gradient \( \nabla\mathcal{L}(\cdot) \), which is crucial for the differential privacy guarantee. The authors specify a family of weight functions \( \alpha(\cdot) \) that ensure the gradient sensitivity is bounded, which is a requisite for the differential privacy mechanisms employed. The authors also define a family of desirable weight functions \( \mathcal{F}(\mathcal{C}, \vec{z}) \) as those that are non-increasing and satisfy:
\begin{equation}
 \forall \vec{x} \in \mathbb{R}^n, \alpha(\|\vec{x} - \vec{z}\|) \leq \frac{c}{2\|\vec{x}-\vec{z}\|_2(\|\vec{x}-\vec{z}\|_2+1)}.
 \end{equation}

\sstitle{Adaptive algorithm for streaming explanation queries}
Patel et al.~\cite{patel2022model} describes an adaptive differentially private algorithm that involves sequentially explaining queries with the aid of differential privacy, using information from previously explained queries to optimize future explanations and manage the privacy budget.
Key insights for this approach include reusing past explanations for similar new queries and ensuring that the initialization of the Differentially Private Gradient Descent (DPGD) is as close as possible to the new query to achieve faster convergence and reduce privacy spending. The authors present a weight function \( \alpha(\|\vec{x} - \vec{z}\|) \), defined as:
\begin{equation}
 \alpha(\|\vec{x} - \vec{z}\|) = \begin{cases} 
1 & \text{if } \|\vec{x} - \vec{z}\| \leq r \\
\frac{c}{2\|\vec{x}-\vec{z}\|_2(\|\vec{x}-\vec{z}\|_2+1)} & \text{else}
\end{cases}
\end{equation}
This weight function is used to identify points similar to \( \vec{z} \) and is employed to ensure stable and consistent local explanations.

Patel et al.~\cite{patel2022model} also introduces the idea of a non-interactive differential privacy mechanism to generate new explanations without additional privacy spending by constructing a proxy dataset from previous explanations.

\subsubsection{Differentially Private Counterfactual Explanations}
Mochaourab et al.~\cite{mochaourab2021robust} develop a differentially private Support Vector Machine (SVM) and introduce methods for generating robust counterfactual explanations. 
Yang et al.~\cite{yang2022differentially} creates a differentially private autoencoder to produce privacy-preserving prototypes for each class label, optimizing perturbations to the input data that minimizes distance to the counterfactual while favoring a specific class outcome.
Hamer et al.~\cite{hamer2023simple} suggests data-driven recourse directions could be privatized, but does not elaborate on providing private multi-step recourse paths. Huang et al.~\cite{huang2023accurate} proposes generating privacy-preserving recourse using a differentially private logistic regression model but does not detail the provision of a multi-step path for recourse.
Pentyala et al.~\cite{pentyala2023privacy} is a pioneer to offer a complete privacy-preserving pipeline that provides counterfactual explanations with differential privacy guarantees. 
Huang et al.~\cite{huang2023accurate} outlines a methodology for incorporating differential privacy (DP) into logistic regression classifiers to offer recourse against membership inference (MI) attacks. Logistic regression is described with weights \( w \) that output a probability score \( f(x) = w^T x = \log \frac{P(y=1|x)}{1-P(y=1|x)} \). The counterfactual distance for instance \( x \) from the target score \( s \) in logistic regression space is given by \( c(x, x') = \frac{s - f(x)}{\| w \|_2^2} \). The decision boundary is set at \( s = 0 \), meaning that \( P(y = 1|x) \) is 0.5 at the threshold. In particular, Huang et al.~\cite{huang2023accurate} introduces two DP methods for recourse generation:

\begin{itemize}
	\item \emph{Differentially Private Model (DPM):} It involves training the logistic regression classifier with DP. An $\epsilon$-DP logistic regression model leads to $\epsilon$-DP counterfactual recourse, using IBM's \emph{diffprivlib} library~\cite{holohan2019diffprivlib} based on Chaudhuri et al.'s mechanism for DP empirical risk minimization~\cite{chaudhuri2011differentially,wang2017differentially}.

\item \emph{Differentially Private Laplace Recourse (LR):} A new method is proposed for DP post-hoc computation of counterfactual recourse that does not touch the underlying logistic regression model training process. It involves:
(1) Applying Laplace noise to the predicted probability score \( Pr'(y = 1|x) = Pr(y = 1|x) + \text{Laplace}(1/\varepsilon) \).
(2) Clamping \( Pr'(y = 1|x) \) to $[0,1]$.
(3) Computing the noisy logistic regression score \( f'(x) \) based on \( Pr'(y = 1|x) \).
(4) Calculating the noisy CFD as \( c'(x, x') = \frac{s - f'(x)}{\| w \|_2^2} \). %
\end{itemize}

Huang et al.~\cite{huang2023accurate} claim that these methods are $\epsilon$-DP. This is explained by starting with applying Laplace noise to the predicted probability and noting that the global sensitivity \( GS_{p(y=1|x)} \) is 1. The process from calculating \( Pr(y = 1|x) \) to \( M_{CFD, Lap}(x) \) is argued to be a post-processing step that retains $\epsilon$-DP, according to the post-processing invariance property of DP~\cite{dwork2014algorithmic}.

Pawelczyk et al.~\cite{pawelczyk2023privacy} proposes that applying DP to a recourse generation algorithm can limit an adversary's balanced accuracy, with a bound expressed as \( BA_A \leq \frac{1}{2} + \frac{1}{2} \cdot e^{-\epsilon} \), where \( \epsilon \) is the privacy loss parameter. However, the authors also acknowledges that while DP offers robust privacy assurances, it is not a fail-safe measure and can significantly reduce accuracy, posing a challenge in maintaining the utility of the explanation.
Pentyala et al.~\cite{pentyala2023privacy} proposes ``PrivRecourse'', a framework for generating privacy-preserving counterfactual explanations.
 The method relies on a two-phase approach: a training phase and an inference phase.
The training phase involves training a differentially private ML model \( f \), clustering the dataset into \( K \) subsets with (\( \epsilon_k, \delta_k \))-DP guarantees, and constructing a graph \( G \) with clusters as nodes~\cite{joshi2023k,lu2020differentially}. Nodes are connected by edges based on distance and density without violating actionable constraints, and the entire graph is published ensuring (\( \epsilon, \delta \))-differential privacy~\cite{abadi2016deep,dwork2014algorithmic}.
During the inference phase, for any query instance \( Z \), a recourse path \( P \) and a counterfactual instance \( Z^* \) that would flip the model's decision to a favorable outcome are computed. This is done by first identifying the nearest node \( Z_1 \) to \( Z \) in \( G \), and then using Dijkstra's algorithm to find the shortest path to the favorable counterfactuals in \( Z_{CF} \)~\cite{wagner2023fast}. 

Hamer et al.~\cite{hamer2023simple} proposes another framework to generate counterfactuals, called the Stepwise Explainable Paths (StEP).
The framework begins by partitioning the dataset \( X \) into \( k \) clusters \( \{X_1, ..., X_k\} \). For a point of interest \( \tilde{x} \), if the model prediction \( f(\tilde{x}) = -1 \) indicating an unfavorable outcome, StEP generates a direction \( \tilde{d}_c \) for each cluster using the formula:
\begin{equation}
\tilde{d}_c = \sum_{x' \in X_c} (x' - \tilde{x})(\alpha(||x' - \tilde{x}||) f(x') = 1)
\end{equation} 
Here, \( \alpha \) is a non-negative function, and \( || \cdot || \) is a rotation invariant distance metric~\cite{sliwinski2019axiomatic}. This process repeats iteratively, with the user updating their point of interest \( \tilde{x} \), until a favourable outcome is achieved.
StEP can be adapted to satisfy (\( \epsilon, \delta \))-differential privacy by adding Gaussian noise to the directions computed. When the distance metric is the \( \ell_2 \) norm, the sensitivity of StEP is upper-bounded by a constant \( C \), and therefore, Gaussian noise with a mean of 0 and standard deviation \( \sigma \geq \frac{C^2 \beta}{\epsilon} \) where \( \beta \geq 2 \log(1.25/\delta) \) can be added to each feature to achieve differential privacy. When multiple directions are provided to a user, and each is (\( \epsilon, \delta \))-differentially private, the overall mechanism is (\( k\epsilon, k^{\delta} \))-differentially private~\cite{dwork2014algorithmic}.

Yang et al.~\cite{yang2022differentially} proposes another DP-based method through the use of a functional mechanism. The functional mechanism does not add noise directly to the optimal parameter set \( w^* \), but to the loss function \( \tilde{L}_D(w) \) by injecting Laplace noises into the coefficients of its polynomial representation. 
The process involves constructing class prototypes in the latent space using a well-trained autoencoder and the functional mechanism through a perturbed training loss. Counterfactual samples are then searched for in the latent space based on these prototypes.
Yang et al.~\cite{yang2022differentially} provides that if the prototype construction process is \( \epsilon \)-differentially private, then the counterfactual explanation process also satisfies DP under the same privacy budget \( \epsilon \). This relies on the post-processing immunity of DP~\cite{dwork2014algorithmic}, which allows for certain noises to be added in the prototype construction process without further affecting subsequent computations.

\subsubsection{DP-Locally Linear Maps}

To create differentially private Locally Linear Maps (LLM), Harder et al.~\cite{harder2020interpretable}  employs the moments accountant technique combined with differentially private stochastic gradient descent (DP-SGD)~\cite{abadi2016deep}. 
The perturbation process involves two main steps per iteration for each minibatch of size \( L \):
(1) Clipping the norm of the datapoint-wise gradient \( h_t(x_n) \) using a threshold \( C \) and adding Gaussian noise to it, resulting in \( \hat{h}_t \):
   $ \hat{h}_t \leftarrow \frac{1}{L} \sum_{n=1}^{L} h_t(x_n) + \mathcal{N}(0, \sigma^2 C^2 I) $.
(2) Updating the LLM parameters in the descending direction:
   $ W_{t+1} \leftarrow W_t - \eta \hat{h}_t $.
This process ensures that the final LLM is \((\epsilon, \delta)\)-differentially private.
To improve the privacy-accuracy trade-off, especially for high-dimensional inputs like images, the author suggest reducing the dimensionality of the parameters by first projecting them onto a lower-dimensional space using a shared matrix \( R_m \), and then perturbing the gradients of the projected parameters~\cite{xue2024revisiting}.

\begin{figure}[!h]
    \centering
    \includegraphics[width=\linewidth]{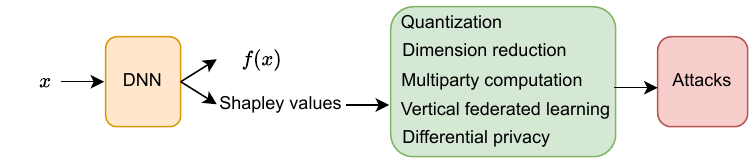}
    \caption{Defences with Privacy-preserving Shapley values.}
    \label{fig:defence-shap}
\end{figure}

\subsection{Defences with Privacy-Preserving SHAP}
Several studies have focused on preserve the privacy of users from explanation using Shapley values, including quatization, dimension reduction, multi-party computation, federated learning, and differential privacy (see~\autoref{fig:defence-shap}).

\sstitle{Quantized Shapley values}
Luo et al.~\cite{luo2022feature} proposes quantization of Shapley values to protect privacy by 
reducing mutual information between input features and their corresponding Shapley values. By restricting the Shapley values to a set number of discrete levels (e.g., 5, 10, or 20 distinct values), the entropy of the Shapley values, \( H(s_i) \), and hence the mutual information \( I(x_i; s_i) \) can be reduced. While quantization has minimal effects on the effectiveness of one attack strategy, it does compromise the accuracy and success rate of another due to the increased range of candidate estimations for a feature, leading to larger estimation errors as per the bounds established earlier. Quantization might also result in two different input samples yielding the same explanation, which is an issue for the privacy-utility balance. 

\sstitle{Low-dimensional Shapley values}
Luo et al.~\cite{luo2022feature} discusses a defensive strategy by suggesting a reduction in the dimensionality of Shapley values. Since the number of Shapley values for a class corresponds to the number of input features, the defence involves only releasing the Shapley values of the top \( k \) features based on their variance, rather than their magnitude. 

\sstitle{Multi-party Shapley values}
Jetchev et al.~\cite{jetchev2023xorshap} introduces secure multiparty computation (MPC), which allows multiple parties to jointly evaluate a public function on their private data without revealing anything other than the function's output. The authors developed a privacy-preserving algorithm, XorSHAP, which operates on top of the Manticore MPC framework. This algorithm is a variant of the TreeSHAP method and retains agnosticism towards the underlying MPC framework. 
The authors discuss the secret sharing of binary decision trees within an MPC setting, where decision trees can be shared secretly and then used in the computation of privacy-preserving algorithms like XorBoost. Jetchev et al.~\cite{jetchev2023xorshap} proves that all subsequent operations and variables in the algorithm are secret and data-independent.

\sstitle{Federated Shapley values}
Wang et al.~\cite{wang2019interpret} discusses interpreting models in the context of Vertical Federated Learning (VFL)~\cite{liu2024dynamic,liu2023long,liu2022privacy,liu2024guaranteeing} where different parties possess different slices of the feature space. Traditional model interpretation methods like Shapley values can reveal sensitive data across parties, making it unsuitable for VFL. To address this, a variant called SHAP Federated is proposed for VFL, particularly for dual-party scenarios involving a host and guest. The host and guest collaboratively develop a machine learning model, with the host owning the label data and part of the feature space, and the guest owning another part. 
The algorithm involves setting values in the instance \( x \) to their original or reference values based on whether a feature is hosted or federated and encrypting IDs when necessary to maintain privacy. Then, predictions are made for each combination of features, and feature importance is calculated from the aggregated prediction results using Shapley values. Features that cannot handle missing values are set to either NA or the median~\cite{lundberg2017unified}.

\sstitle{Differentially Private Shapley values}
Luo et al.~\cite{luo2022feature} points out that DP is not suitable for local interpretability methods. For DP to be effective, the explanations for any two different private samples must be indistinguishable, which would reduce the utility of Shapley values as they would become too similar across different samples. As a result, DP cannot be applied to the current problem of maintaining interpretability while defending against attacks that leverage Shapley values.

Watson et al.~\cite{watson2022differentially} discusses the computational challenges of calculating Shapley values due to their expensive nature and the privacy concerns in using large portions of datasets for each query. The authors introduce an estimation algorithm that utilizes only a small fraction of data, taking advantage of the property that larger datasets reduce the marginal contributions of individual data points, which are proportionally smaller. 
The algorithm is shown to satisfy \( \epsilon \)-differential privacy with a coalition sample complexity of \( O(\ln(n)) \)~\cite{watson2022differentially}.
Watson et al.~\cite{watson2022differentially} emphasises the cost advantages of the Layered Shapley approach, which uses fewer data points and has lower computational and data access costs, offering privacy benefits. 

\begin{figure}[!h]
    \centering
    \includegraphics[width=.95\linewidth]{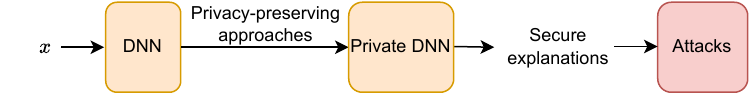}
    \caption{Defences with privacy-preserving ML models.}
    \label{fig:defence-model}
\end{figure}

\subsection{Defences with Privacy-preserving ML models}
\begin{figure}[!h]
    \centering
    \includegraphics[width=.9\linewidth]{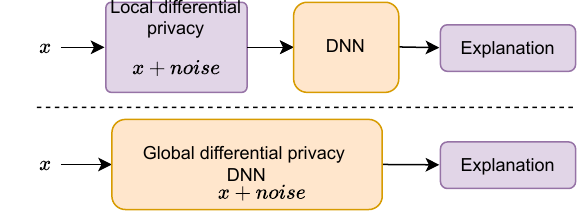}
    \caption{Local and global differential privacy schemes proposed in~\cite{naidu2021differential}.}
    \label{fig:defense-differential-local-global}
\end{figure}

To protect user privacy, privacy-preserving ML models have been trained to resist against attacks (see~\autoref{fig:defence-model}).
Naidu et al.~\cite{naidu2021differential} discusses two primary models of implementing differential privacy: Local DP, where noise is added directly to user data before it is shared, ensuring data privacy against untrusted parties; and Global DP, where a trusted central entity applies differentially private algorithms like DP-SGD~\cite{abadi2016deep} to the collected data to produce models or analyses with limited information leakage (see \autoref{fig:defense-differential-local-global}). Interpreting models trained with differential privacy is challenging due to the noise added during training, which obfuscates the model's decision-making process~\cite{patel2022model}. Naidu et al.~\cite{naidu2021differential} investigates the interpretability of differentially private models by establishing the first benchmark for interpretability in deep neural networks (DNNs) trained with differential privacy. 

Liu et al.~\cite{liu2024please} develops a model-level defense by employing Differentially-Private Stochastic Gradient Descent (DP-SGD)~\cite{bu2023differentially}, to build inherently private models. The process involves automatic configuration of gradient clipping and the selection of `MixOpt' as the clipping model, uniformly applied across all model layers. While DP-SGD can reduce the effectiveness of membership inference attacks, it also significantly decreases classification accuracy, even with a large epsilon $\epsilon$. 
Findings indicate that 
attribution maps become less informative than even methods not considering model parameters~\cite{hooker2019benchmark}. This underscores the challenge of balancing between defense capability and performance utility, as effective defense mechanisms like DP-SGD can significantly impact model accuracy and the quality of explanations provided.

Mochaourab et al.~\cite{mochaourab2021robust} outlines a method for providing differential privacy to SVM classifiers by perturbing the optimal weight vector \( w^* \) with additive Laplace noise. 
The perturbed weight vector \( \tilde{w} \) is given by \( \tilde{w} := w^* + \mu \), where \( \mu \) consists of i.i.d. Laplace random variables \( \mu_i \sim \text{Lap}(0, \lambda) \). This perturbation ensures \( \beta \)-differential privacy for \( \lambda \geq 4C_k\sqrt{F}/(\beta n) \), with certain conditions on the kernel function \( \phi \).
Mochaourab et al.~\cite{mochaourab2021robust} introduces robust counterfactual explanations for SVM classifiers, providing explanations for classification results that account for the uncertainty introduced by the differential privacy mechanism. 
For the optimization problem, a root of the function \( g \), defined as: 
\begin{equation}
 y'f_{\phi}(x, \tilde{w}) - \lambda\sqrt{2\ln(2/(1-p))}\|\phi(x)\| \leq 0
 \end{equation}
 is considered as a robust counterfactual explanation. Efficient solutions to this optimization problem are proposed using convex optimization solvers like CVXPY for linear SVM or a bisection method for non-linear SVM.
The solution implies that a domain expert's input is required to determine prototypes representing each class when direct access to test data is not available due to privacy considerations. 
A bisection method used for finding robust counterfactual explanations in non-linear SVMs is also developed~\cite{mochaourab2021robust}.

Veugen et al.~\cite{veugen2022privacy} uses local foil trees to explain the decisions of a black-box model without accessing its training data. By generating synthetic data points that are close to the user's data point, classifying them through the model, and then training a decision tree in a secure manner, the method constructs explanations in terms of feature thresholds~\cite{van2018contrastive}. 
This process utilises secret-shared data and secure multi-party computation~\cite{lindell2020secure} to ensure that no sensitive information from the model or its training data is disclosed, except for the minimal necessary details required to provide the user with an explanation for the classification outcome.

\subsection{Defences with Perturbations}

Jia et al.~\cite{jia2019memguard} introduces a defence technique called MemGuard, differing from other strategies that modify the training process. MemGuard cleverly injects perturbations into the confidence scores produced by the model for each input, transforming these altered scores into adversarial examples aimed at misleading attack models. 
However, the primary limitation of MemGuard is its focus on distorting the model's output by adding noise, which does not protect the attribution maps, thus failing to completely deter the attacks~\cite{liu2024please}.

\begin{figure}[!h]
    \centering
    \includegraphics[width=\linewidth]{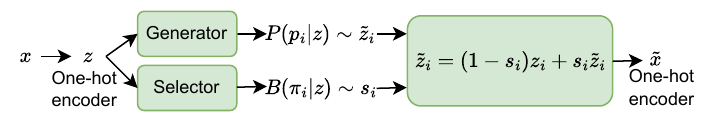}
    \caption{The approach of generating diverse counterfactuals to reduce the privacy risk in re-identification~\cite{vo2023feature}.}
    \label{fig:defence-diverse-counterfactual}
\end{figure}

Vo et al.~\cite{vo2023feature} describes a methodology for addressing the trade-off between diversity and sparsity in the features modified to form a counterfactual. As shown in \autoref{fig:defence-diverse-counterfactual}, it introduces a local feature-based perturbation distribution \(P(\tilde{z}_i | z)\) for each mutable feature \(z_i\), along with a selection distribution \( \text{Bernoulli}(\pi_i | z) \) to control sparsity. To form a counterfactual example \( \tilde{z} \), the method samples from these distributions and updates mutable features, maintaining validity by maximising the likelihood of the counterfactuals to alter the original outcome.

Olatunji et al.~\cite{OlatunjiRFK23} discusses a defence mechanism for feature-based explanations.
It involves perturbing each explanation bit, where an explanation is represented as a bit mask, by using a randomised response mechanism. The perturbation probability for flipping each bit \( \mathcal{E}_{xi} \) is determined by a privacy budget \( \epsilon \):
\[ \text{Pr}(\mathcal{E}_{xi}' = 1) = \begin{cases} 
\frac{e^\epsilon}{e^\epsilon+1} & \text{if } \mathcal{E}_{xi} = 1, \\
\frac{1}{e^\epsilon+1} & \text{if } \mathcal{E}_{xi} = 0, 
\end{cases} \]
where \( \mathcal{E}_{xi} \) and \( \mathcal{E}_{xi}' \) are the true and perturbed \( i^{th} \) bits of explanation, respectively. This method ensures \( d\epsilon \)-local differential privacy for an explanation with \( d \) dimensions.

\subsection{Defences with Anonymisation}

\sstitle{k-Anonymity}
Goethals et al.~\cite{goethals2023privacy} presents a unique application of k-anonymity aimed at ensuring anonymity within counterfactual explanations, as opposed to anonymising an entire dataset. This approach is particularly relevant when the dataset is not intended to be fully public. The authors define a counterfactual instance as k-anonymous if its quasi-identifiers -- the partially identifying attributes -- could apply to at least k individuals within the training set. In turn, a counterfactual explanation inherits this k-anonymity if it is derived from such a k-anonymous instance.
However, while counterfactual explanations usually aim to change the outcome of a model's prediction, k-anonymous counterfactuals can include a range of instances beyond those used to generate the explanation, leading to uncertainty about whether all values in this range would lead to a change in the prediction.

\begin{figure}[!h]
    \centering
    \begin{subfigure}[b]{\linewidth}
        \centering
        \includegraphics[width=\linewidth]{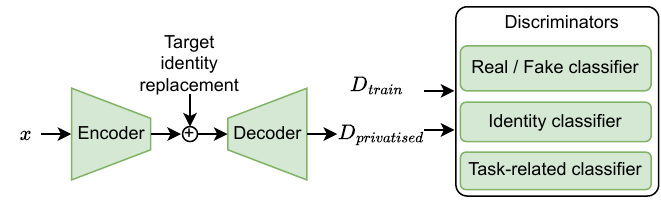}
        \caption{The PPRL-VGAN model proposed in~\cite{chen2018vgan}.}
        \label{fig:defence-anonymisation-PPRL-VGAN}
    \end{subfigure}
    \begin{subfigure}[b]{\linewidth}
        \centering
        \includegraphics[width=\linewidth]{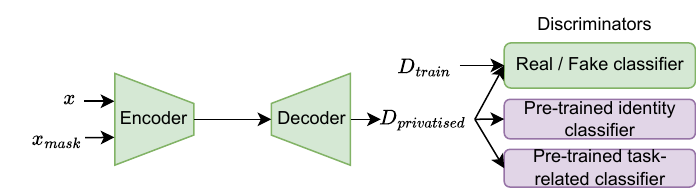}
        \caption{The WGAN-GP framework using pre-trained identifier and task-related classifier~\cite{montenegro2021privacy}.}
        \label{fig:defence-anonymisation-WGAN-GP}
    \end{subfigure}
    \begin{subfigure}[b]{\linewidth}
        \centering
        \includegraphics[width=\linewidth]{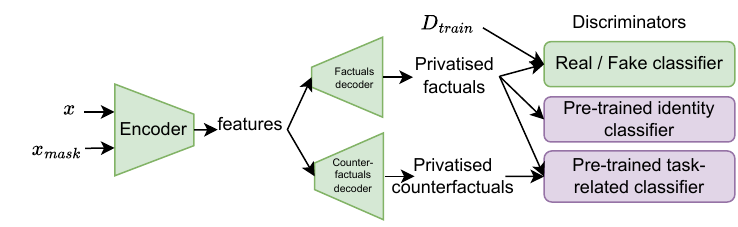}
        \caption{Privatised counterfactual samples are generated by a counterfactual decoder~\cite{montenegro2021privacy}.}
        \label{fig:defence-anonymisation-counterfactual}
    \end{subfigure}    
    \caption{Three approaches to defend against attacks with anonymisation. Subfigures (a-b) focus on generating privatised factual samples, while (c) aims to generate privatised counterfactual samples. $x_{mask}$ is the mask data, and $D_{privatised}$ is the generated privatised data.}
    \label{fig:defence-anonymisation}
\end{figure}

\sstitle{Privatised Factual Samples}
Montenegro et al.~\cite{montenegro2022privacy} 
argues that an explanation should not reveal sensitive personal identity information while remaining realistic and informative regarding the decision-making process. Montenegro et al.~\cite{montenegro2022privacy} outlines an optimisation objective which involves minimising three loss functions, one for privacy, one for realism, and one for explanatory evidence, each weighted by a non-negative parameter. The distance between a privatised image and the source image is minimised, ensuring that the privatised image is sufficiently different from any identity in the training data to preserve anonymity~\cite{montenegro2021privacy}.

Montenegro et al.~\cite{montenegro2021privacy} develops a privacy-preserving network with multi-class identity recognition designed for case-based explanations. The network seeks to preserve privacy by promoting a uniform distribution across identities, making identity recognition akin to random guessing. 
The PPRL-VGAN model~\cite{chen2018vgan} (see~\autoref{fig:defence-anonymisation-PPRL-VGAN}), which intentionally collapses to the replacement identity and task-related class, is replaced with a WGAN-GP framework that uses a Wasserstein loss with a gradient penalty to stabilise the discriminator (see~\autoref{fig:defence-anonymisation-WGAN-GP}). This change, alongside using interpretability saliency maps for reconstruction of relevant task-related features, aims to retain the explanatory value in the privatised images~\cite{montavon2017explaining}. 
Montenegro et al.~\cite{montenegro2021privacy} also introduces another privacy-preserving network that utilises a Siamese identity recognition framework to enhance privacy in domains with scarce images per subject. 
They employ a contrastive loss function for training, defined as \( \text{ContrastiveLoss} = \frac{1}{2} \times Y \times ED^2 + \frac{1}{2} \times (1 - Y) \times [\max(0, m - ED)]^2 \), where \( Y \) is the label indicating if the image pair is of the same identity, \( ED \) is the Euclidean distance between embeddings, and \( m \) is a margin. The Siamese network ensures the privatised image is distinct in identity from the original and others in the dataset.

\sstitle{Privatised Counterfactual Samples}
Montenegro et al.~\cite{montenegro2021privacy} also generates counterfactual explanations from the privatised samples. As shown in~\autoref{fig:defence-anonymisation-counterfactual}, a counterfactual generation module, in the form of a decoder, is added to the above privacy-preserving network to map an image's latent representation to its counterfactual. This decoder is designed to make minimal alterations to the privatised factual explanations to change their predicted class, thereby minimising the pixel-wise distance between the factual and counterfactual explanations while altering the image's task-related prediction. Saliency masks and explanatory features are used to guide changes to image regions that are relevant to the explanation. 
The loss function for the counterfactual decoder training is represented as \( L_C = E_{I,M\sim p_{data}}[ \lambda_{x}[F(I) \times (1 - M) - C(I) \times (1 - M)]^2 + \lambda_{D}Exp(D_{exp}(I) \times \log(1 - D_{exp}(C(I))) ] \), where \( F(I) \) and \( C(I) \) denote the privatized factual and counterfactual explanations, respectively, and \( \lambda_{x} \) and \( \lambda_{D} \) are weights controlling the importance of each term in the loss function.

\subsection{Defences with Collaborative Explanation}

Domingo et al.~\cite{domingo2019collaborative} presents methods for collaborative rule-based model approximation without the direct use of a model simulator. It suggests that users can employ simulators to interact with a concealed model to obtain responses for certain feature sets, which although limited and controlled, can help deduce how the model makes decisions. While simulators prevent full transparency of the model and often limit the number of queries to prevent misuse, users can collaborate by querying the model for various feature sets and publishing the predictions. This collective data can then be mined for decision rules to approximate the model's logic.

\subsection{Defences against Reconstruction Attacks}

\begin{figure}[!h]
    \centering
    \includegraphics[width=\linewidth]{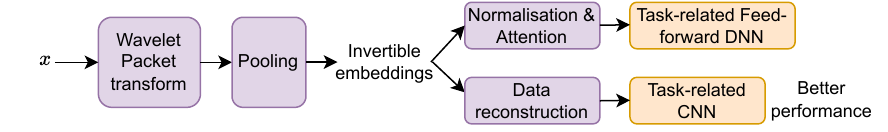}
    \caption{The ``DeepFixCx'' model uses compression techniques (\ie wavelet packet transform and a pooling function) for preserving privacy and explicability~\cite{gaudio2023deepfixcx}.}
    \label{fig:defence-compression}
\end{figure}

Gaudio et al.~\cite{gaudio2023deepfixcx} proposes the ``DeepFixCx'' model, an approach that utilises wavelet packet transforms and spatial pooling for image compression that preserves privacy and explicability (see~\autoref{fig:defence-compression}). The method relies on analysing images with multi-scale wavelet-based methods, allowing local regions of pixels to be summarised at multiple scales. 
The wavelet packet transform offers several benefits, such as facilitating image processing with deep learning libraries, ensuring that all coefficient values represent equally-sized pixel regions, and maintaining consistency with boundary effects. 
``DeepFixCx'' provides a trade-off between compressing images for efficiency while still retaining enough detail for reconstruction and privacy preservation. 
Gaudio et al.~\cite{gaudio2023deepfixcx} also outlines methods for inverse wavelet packet transform for image reconstruction, which can restore images from compressed representations to their original size. This model offers a privacy-conscious method to process images for various applications, including medical imaging, by removing local spatial information, allowing for the preservation of privacy without the need for additional learning.

\section{Published Resources}
\label{sec:resource}

\begin{table*}[!h]
    \centering
    \caption{Published Algorithms and Models} %
    \label{tab:algorithms}
    \begin{adjustbox}{max width=\textwidth}
    \begin{threeparttable}
    \begin{tabular}{l|c|c|c|c|l}
        \toprule
	\textbf{Algorithms} & \textbf{Year} & \textbf{Target Explanations} & \textbf{Attacks} & \textbf{Defenses} &  \textbf{Code Repository} \\
	\midrule
	L2C~\cite{vo2023feature} & 2023 & Counterfactual & -- & Perturbation &\rurl{github.com/isVy08/L2C/} \\
	GSEF~\cite{OlatunjiRFK23} & 2023 & Feature-based & Graph Extraction & Perturbation & \rurl{github.com/iyempissy/graph-stealing-attacks-with-explanation} \\
	Ferry et al.~\cite{ferry2023probabilistic} & 2023 & Interpretable models & Data Reconstruction & - & \rurl{github.com/ferryjul/ProbabilisticDatasetsReconstruction} \\
	DeepFixCX~\cite{gaudio2023deepfixcx} & 2023 & Case-based & Identity recognition & Anonymisation & \rurl{github.com/adgaudio/DeepFixCX} \\
	DP-XAI & 2023 & ALE plot & - & Differential Privacy & \rurl{github.com/lange-martin/dp-global-xai} \\
	Duddu et al.~\cite{duddu2022inferring} & 2022 & Gradient/Perturbation-based & Attribute Inference & - & \rurl{github.com/vasishtduddu/AttInfExplanations} \\
	DataShapley~\cite{watson2022differentially} & 2022 & Shapley & - & Differential Privacy & \rurl{github.com/amiratag/DataShapley} \\
	MEGEX~\cite{miura2021megex} & 2021 & Gradient-based & Model Extraction & - & \rurl{github.com/cake-lab/datafree-model-extraction} \\
	Mochaourab et al.~\cite{mochaourab2021robust} & 2021 & Counterfactual & - &  Private SVM & \rurl{github.com/rami-mochaourab/robust-explanation-SVM} \\
	Gillenwater et al.~\cite{gillenwater2021differentially} & 2021 & Quantiles & - & Differential Privacy & \rurl{github.com/google-research/google-research/tree/master/dp\_multiq} \\
	DP-LLM~\cite{harder2020interpretable} & 2020 & Locally linear maps & - & Differential Privacy & \rurl{github.com/frhrdr/dp-llm} \\
	MRCE~\cite{aivodji2020model} & 2020 & Counterfactual & Model Extraction & - & \rurl{github.com/aivodji/mrce} \\
Federated SHAP~\cite{wang2019interpret} & 2019 &  Shapley & - & Federated & \rurl{github.com/crownpku/federated_shap} \\
         \bottomrule
    \end{tabular}
    \end{threeparttable}
    \end{adjustbox}
\end{table*}

\subsection{Published Algorithms}

Several algorithm and model implementations have been pivotal to foundational experiments in maintaining privacy within model explanations.
\autoref{tab:algorithms} provides a consolidated list of published algorithms and models, categorised by their release year (ranging from 2019 to 2023), the types of explanations they target (such as Counterfactual, ALE plot, Shapley values), potential attacks (like Perturbation, Graph Extraction), and corresponding defences (including Differential Privacy, Anonymisation). Each listed algorithm, such as L2C, DP-XAI, and GSF, among others, is accompanied by a link to its code repository on GitHub, allowing for easy access to their implementation details for further exploration or usage.

\subsection{Published Datasets}

\begin{table*}[!h]
    \centering
    \caption{Highlighted Datasets} %
    \label{tab:datasets}
    \begin{adjustbox}{max width=\textwidth}
    \begin{threeparttable}
    \begin{tabular}{c|c|c|c|c|c|l}
        \toprule
	\textbf{Category} & \textbf{Dataset} &  \textbf{\#Items} &  \textbf{Disk Size} & \textbf{Downstream Explanations} & \textbf{Experimented in} &  \textbf{URL} \\
	\midrule
	Image & MNIST & 70K & 11MB &  Counterfactuals, Gradient & \cite{huang2023accurate,yang2022differentially,zhao2021exploiting,milli2019model} & \rurl{www.kaggle.com/datasets/hojjatk/mnist-dataset} \\
	& CIFAR & 60K & 163MB & Gradient & \cite{miura2021megex,shokri2021privacy,milli2019model,liu2024please} & \rurl{www.cs.toronto.edu/~kriz/cifar.html} \\
	& SVHN & 600K & 400MB+ & Gradient & \cite{miura2021megex} & \rurl{ufldl.stanford.edu/housenumbers/} \\
	 & Food101 & 100K+ & 10GB &  Case-based & \cite{gaudio2023deepfixcx} & \rurl{www.kaggle.com/datasets/dansbecker/food-101} \\
	 & Flowers102 & 8K+ & 300MB+ &  Case-based & \cite{gaudio2023deepfixcx} & \rurl{www.robots.ox.ac.uk/~vgg/data/flowers/102/} \\
	 & Cervical & 8K+ & 46GB+ &  Case-based, Interpretable Models & \cite{gaudio2023deepfixcx} & \rurl{www.kaggle.com/competitions/intel-mobileodt-cervical-cancer-screening} \\
	 & CheXpert & 220K+ & GBs &  Case-based, Interpretable Models & \cite{gaudio2023deepfixcx} & \rurl{stanfordmlgroup.github.io/competitions/chexpert/} \\
	 & Facial Expression & 12K+ & 63MB & Black-box & \cite{patel2022model} & \rurl{www.kaggle.com/datasets/msambare/fer2013} \\
	 & Celeb & 200K & GBs & Gradient & \cite{zhao2021exploiting} & \rurl{mmlab.ie.cuhk.edu.hk/projects/CelebA.html} \\
	\hline
	Tabular & Adult & 48K+ & 10MB & Counterfactuals, Shapley, Gradient, Perturbation  & 10+ (\cite{huang2023accurate,ferry2023probabilistic,pentyala2023privacy} etc.) & \rurl{archive.ics.uci.edu/ml/datasets/adult} \\
	 & COMPAS & 7K+ & 25MB & Counterfactuals, Shapley, Gradient, Perturbation  & \cite{ferry2023probabilistic,duddu2022inferring} & \rurl{www.kaggle.com/datasets/danofer/compass} \\
  & FICO &  10K+ & $\leq $ 1MB & Counterfactuals, Shapley  & \cite{huang2023accurate,wang2022dualcf,pentyala2023privacy,pawelczyk2023privacy} & \rurl{community.fico.com/s/explainable-machine-learning-challenge} \\
   & Boston Housing &  500+ & $\leq $ 1MB & Counterfactuals, Shapley  & \cite{wang2022dualcf} & \rurl{www.kaggle.com/code/prasadperera/the-boston-housing-dataset} \\
   & German Credit & 1K & $\leq $ 1MB & Counterfactuals, Shapley, Gradient, Perturbation & \cite{vo2023feature,goethals2023privacy,yang2022differentially,duddu2022inferring} & \rurl{archive.ics.uci.edu/dataset/144/statlog+german+credit+data} \\
   & Student Admission & 500 & $\leq $ 1MB & Counterfactuals, Shapley & \cite{vo2023feature} & \rurl{www.kaggle.com/datasets/mohansacharya/graduate-admissions} \\
   & Student Performance & 10K & $\leq $ 1MB & Counterfactuals, Shapley & \cite{vo2023feature} & \rurl{www.kaggle.com/datasets/nikhil7280/student-performance-multiple-linear-regression} \\
   & GMSC &  150K+ & 15MB & Counterfactuals, Shapley  & \cite{wang2022dualcf,naretto2022evaluating} & \rurl{www.kaggle.com/c/GiveMeSomeCredit/data} \\
   & Diabetes & 100K+ & 20MB & Counterfactuals, Shapley & \cite{pawelczyk2023privacy,luo2022feature,yang2022differentially,watson2022differentially,shokri2021privacy} & \rurl{archive.ics.uci.edu/dataset/296/diabetes+130-us+hospitals+for+years+1999-2008} \\
   & Breast Cancer & 569 & $<1MB$ & Interpretable models, Counterfactuals & \cite{mochaourab2021robust} & \rurl{archive.ics.uci.edu/ml/datasets/breast+cancer} \\
   \hline
   Graph & Cora & 2K+ & 4.5MB  & Feature-based  & \cite{OlatunjiRFK23} & \rurl{relational.fit.cvut.cz/dataset/CORA} \\
    & Bitcoin & 30K  & $\leq $ 1MB  & Feature-based  & \cite{OlatunjiRFK23} & \rurl{snap.stanford.edu/data/soc-sign-bitcoin-alpha.html} \\
    & CIC-IDS2017 & 2.8M+ & 500MB &  Counterfactuals & \cite{kuppa2021adversarial} & \rurl{www.unb.ca/cic/datasets/ids-2017.html} \\
     \hline
	 Text & IMDB Review & 50K & 66MB  & Black-box  & \cite{patel2022model} & \rurl{ai.stanford.edu/~amaas/data/sentiment/} \\
         \bottomrule
    \end{tabular}
    \end{threeparttable}
    \end{adjustbox}
\end{table*}

The datasets most commonly utilized for privacy-preserving model explanations are depicted in \autoref{tab:datasets}. We categorize these datasets into various groups based on their application domains. Important datasets are described below.

\sstitle{Image}
\textit{The CIFAR dataset}~\cite{CIFAR} consists of two parts. The initial subset, CIFAR-10, comprises ten categories of objects, each with six thousand images. These categories include airplanes, automobiles, various animals, and trucks. The training set consists of five thousand randomly selected images per category, with the remaining images used as test examples. The second section, CIFAR-100, contains 600 images for each of its 100 classes. These classes are further grouped into 20 superclasses, each containing five classes.

\textit{The SVHN dataset}~\cite{SVHN} was compiled using automated methods and Amazon Mechanical Turk from an extensive collection of Google Street View images. It encompasses nearly 600,000 labeled characters, comprising complete numbers and chopped digits in a 32x32 pixel format similar to MNIST. It consists of three subsets: over seventy thousand samples for training, twenty thousand for testing, and approximately half a million additional samples.

\textit{The Food101 dataset}~\cite{bossard2014food} was created by gathering images from foodspotting.com, including 101 popular dishes with 750 training and 250 test images per class. Training images were intentionally left uncleaned to simulate real-world noise. All images were resized, resulting in a total of 101,000 diverse food images.

\sstitle{Text}
\emph{The IMDB/Amazon movie reviews dataset}~\cite{ni2019justifying} contains 8,765,568 movie reviews sourced from the Amazon review dataset, along with an additional 50,000 reviews from the IMDB large review dataset. These reviews are represented as binary vectors using the top 500 words. Each review is classified as either positive (+1) or negative (-1).

\sstitle{Tabular}
\textit{The UCI Adult Income dataset}~\cite{ferry2023probabilistic} provides insights from the 1994 U.S. census, aiming to forecast whether an individual earns over \$50,000 annually. Numeric features are divided into quantiles, while categorical features are transformed into binary form through one-hot encoding. This dataset comprises 48,842 examples, each characterized by 24 binary features.

\textit{The Diabetes dataset}~\cite{strack2014impact} contains information from diabetic patients gathered via two methods: traditional paper records and an automated recording system. While paper records indicate time slots of the day, the automated system timestamps occurrences accurately. Each entry in the dataset comprises four fields separated by tabs, with records separated by new lines.

\textit{FICO Explainable Machine Learning Challenge:}
The dataset contains anonymized HELOC (Home Equity Line of Credit) applications from homeowners~\cite{sokol2019counterfactual,huang2023accurate}. HELOCs are credit lines that banks offer based on a percentage of a home's equity. Applicants in the dataset have requested credit lines ranging from \$5,000 to \$150,000. 
The prediction task is to determine the binary target variable ``RiskPerformance'', where ``Bad'' signifies a 90-day overdue payment at least once in 24 months, and ``Good'' indicates timely payments without significant delinquency.

\sstitle{Graph}
\textit{Cora}~\cite{sen2008collective} is a dataset focused on citations, where each node represents a research article. If one article cites another, there's an edge between them. Each node is labeled with its article category. The features of each node are represented by a binary word vector, indicating whether a word is present or absent in the article's abstract.

\textit{The Bitcoin dataset}~\cite{kumar2016edge} is a network representation of trading accounts within the Bitcoin ecosystem. In this dataset, each trading account is depicted as a node, and there are weighted edges connecting pairs of accounts, symbolizing the level of trust between them. The weights range from +10, indicating complete trust, to -10, signifying complete distrust. Each node is labeled to denote its trustworthiness status. The feature vector associated with each node is derived from ratings provided by other users, including metrics such as average positive or negative ratings. 

\textit{The CICIDS17 dataset}, collected under controlled conditions, contains network traffic data in both packet-based and bidirectional flow-based formats. Each flow in the dataset is associated with over 80 features, capturing various aspects of network behavior. 
The dataset is organized into eight groups of features extracted from raw pcaps, including interarrival times, active-idle times, flags-based features, flow characteristics, packet counts with flags, and average bytes and packets sent in various contexts.

\subsection{Evaluation Metrics}

\autoref{tab:metrics} provides the formulas and usages for common metrics in privacy attacks and defences on model explanations. We summarize their descriptions below.

\begin{table*}[!h]
    \centering
    \footnotesize
    \caption{Highlighted Evaluation Metrics}
    \label{tab:metrics}
    \begin{adjustbox}{max width=\textwidth}
    \begin{threeparttable}
    \begin{tabular}{c|p{2.5cm}|p{5cm}|p{7.5cm}}
        \toprule
	\textbf{Category} & \textbf{Evaluation Metrics} & \textbf{Formula/Description} &  \textbf{Usage}\\
	\midrule
Explanation Utility & Counterfactual validity~\cite{goethals2023privacy} & $\text{Pureness} = \frac{\text{\# value combinations with desired outcome}}{\text{\# value combinations}}$ & Assess the range of attribute values within k-anonymous counterfactual instances. Consider all attributes, including those beyond quasi-identifiers \\
\cline{2-4}
& Classification metric~\cite{goethals2023privacy} & $ CM = \frac{\sum_{i=1}^{N} \text{penalty}(tuple_i)}{N} $ & Assess equivalence classes within anonymized datasets, focusing on class label uniformity. \\
\cline{2-4}
 & Faithfulness (RDT-Fidelity)~\cite{OlatunjiRFK23,funke2022z} & $ \mathcal{F}(\mathcal{E}_X) = \mathbb{E}_{Y_{\mathcal{E}_X} | Z \sim \mathcal{N}} \left[ 1_{f(X)=f(Y_{\mathcal{E}_X})} \right] $ & Reflect how often the model's predictions are unchanged despite perturbations to the input, which would suggest that the explanation is effectively capturing the reasoning behind the model's predictions. \\
 \cline{2-4}
 & Sparsity~\cite{OlatunjiRFK23,funke2022z} & $ H(p) = -\sum_{f \in M} p(f) \log p(f) $ & A complete and faithful explanation to the model should inherently be sparse, focusing only on a select subset of features that are most predictive of the model's decision. \\
 \hline
 Information Loss & Normalised Certainty Penalty (NCP)~\cite{goethals2023privacy} & $ \text{NCP}(G) = \sum_{i=1}^{d} w_i \cdot \text{NCP}_{A_i}(G) $ & Higher NCP values indicate a greater degree of generalization and more information loss. This metric helps in assessing the balance between data privacy and utility. \\
 \cline{2-4}
 & Discernibility~\cite{goethals2023privacy} & $ C_{DM}(g, k) = \sum_{VE \,s.t.\, |E| \geq k} |E|^2 + \sum_{VE \,s.t.\, |E| < k} |D||E| $& Measure the penalties on tuples in a dataset after k-anonymization, reflecting how indistinguishable they are post-anonymization \\
 \cline{2-4}
 & Approximation Loss~\cite{goethals2023privacy} & $ \mathcal{E}(\hat{\phi}, \mathcal{Z}, f(X)) \triangleq \mathbb{E} [\mathcal{L}(\hat{\phi}, \mathcal{Z}, f(X)) - \mathcal{L}(\phi^*, \mathcal{Z}, f(X))]. $& Measure the error caused by randomness added when minimizing the privacy loss as the expected deviation of the randomized explanation from the best local approximation \\
 \cline{2-4}
 & Explanation Intersection~\cite{OlatunjiRFK23,funke2022z} & The percentage of bits in the original explanation that is retained in the privatised explanation after using differential privacy & The higher the better but due to privacy-utility trade-off, this metric should not be 100\%.
\\
 \hline
 Privacy Degree & $k$-anonymity~\cite{goethals2023privacy} & A person's information is indistinguishable from at least k-1 other individuals. & Refers to the number of individuals in the training dataset to whom a given explanation could potentially be linked~\cite{goethals2023privacy}.\\
 \cline{2-4}
 & Information Leakage~\cite{patel2022model} & $ Pr_{i=1..k}\hat{\phi}(\mathbf{z_i}, X, f_D(X)) \leq e^{\hat{\varepsilon}} \cdot Pr[\hat{\phi}(\mathbf{z_i}, X, f'_D(X)) : \forall i] + \hat{\delta} $ & If an adversary can access model explanations, they would not gain any additional information that could help in inferring something about the training data beyond what could be learned from the model predictions alone \\
 \cline{2-4}
 & Privacy Budget & The total privacy budget for all queries is fixed at (\( \varepsilon, \delta \)).  & The explanation algorithm must not exceed the overall budget across all queries. Stricter requirement (\( \varepsilon_{min}, \delta_{min} \)) is set for each individual query.
 \\
\hline
Attack Success & Precision/Recall/F1~\cite{duddu2022inferring} & $Prec = \frac{TP}{TP+FP}$,  $Rec = \frac{TP}{TP+FN}$, $F1 = 2 \times \frac{\text{precision} \times \text{recall}}{\text{precision} + \text{recall}}$ & Evaluate an attack's effectiveness in correctly and completely identifying the properties it is designed to infer.
 \\
 \cline{2-4}
 & Balanced Accuracy~\cite{liu2024please,pawelczyk2023privacy,huang2023accurate} & $
BA = \frac{TPR + TNR}{2}
$ & Measures the accuracy of attack (e.g. membership prediction in membership inference attacks), on a balanced dataset of members and non-members. 
 \\
 \cline{2-4}
 & ROC/AUC~\cite{huang2023accurate,pawelczyk2023privacy,liu2024please,ferry2023probabilistic,OlatunjiRFK23} & The ROC curve plots the true positive rate against the false positive rate at various threshold settings. & An AUC near 1 indicates a highly successful privacy attack, while an AUC close to 0.5 suggests no better performance than random guessing.
 \\
 \cline{2-4}
 & TPR at Low FPR~\cite{liu2024please,huang2023accurate,pawelczyk2023privacy} & Report TPR at a fixed FPR (e.g., 0.1\%). & If an attack can pinpoint even a minuscule fraction of the training dataset with high precision, then the attack ought to be deemed effective.\\
 \cline{2-4}
 & Mean Absolute Error (MAE)~\cite{luo2022feature} & $\ell_1 (\hat{x}, x) = \frac{1}{mn} \sum_{j=1}^{m} \sum_{i=1}^{n} | \hat{x}_i^j - x_i^j |,$ & Gives an overview of how accurately an attack can reconstruct private inputs by averaging the absolute differences across all samples and features. \\
 \cline{2-4}
 & Success Rate (SR)~\cite{luo2022feature} & $ SR = \frac{|\hat{X}_{val} \neq \perp|}{mn}$ & The ratio of successfully reconstructed features to the total number of features across all samples \\
 \cline{2-4}
  & Model Agreement~\cite{wang2022dualcf} & $\text{Agreement} = \frac{1}{n} \sum_{i=1}^{n} 1_{f_\theta(x_i) = h_\phi(x_i)}.$ & A higher agreement indicates that the substitute model is more similar to the original model. When comparing two model extraction methods with the same agreement, the one with the lower standard deviation is preferred. \\
  \cline{2-4}
  & Average Uncertainty Reduction~\cite{ferry2023probabilistic} & $Dist(\mathcal{D}^M, \mathcal{D}^{Orig}) = \frac{1}{n \cdot d} \sum_{i=1}^{n} \sum_{k=1}^{d} \frac{H(\mathcal{D}^M_{i,k})}{H(\mathcal{D}_{i,k})}$ & The degree to which a data reconstruction attack is accurate, measured by the reduction in uncertainty across all features of all samples in the dataset \\
	\bottomrule
    \end{tabular}
    \end{threeparttable}
    \end{adjustbox}
\end{table*}

\subsubsection{Explanation utility}
Protecting the privacy might reduce the utility of explanations. Several metrics have been proposed to measure the utility of explanations after privacy protection.

\sstitle{Counterfactual validity}
Goethals et al.~\cite{goethals2023privacy} proposes a pureness metric to measure the validity of counterfactual explanations. It involves assessing the range of attribute values within k-anonymous counterfactual instances. It is important to consider all attributes, including those beyond quasi-identifiers. For categorical attributes, the focus is on the values within the k-anonymous instance, whereas for numerical attributes, the consideration extends to those values also present in the training set.
The \emph{pureness} of a k-anonymous counterfactual explanation is defined by the formula:
\[ \text{Pureness} = \frac{\# \text{ value combinations with desired outcome}}{\# \text{ value combinations}} \]
Practically, it is approximated by querying the model with a set number of random combinations (e.g., 100) to see how many result in the desired prediction outcome. Pureness represents the proportion of these combinations that lead to the desired outcome, aiming for as high a percentage as possible, ideally 100\%.

\sstitle{Classification metric}
The classification metric (CM) is used to assess equivalence classes within anonymised datasets, focusing on class label uniformity~\cite{goethals2023privacy}. It is calculated as:
\[ CM = \frac{\sum_{i=1}^{N} \text{penalty}(tuple_i)}{N} \]
Here, \( N \) is the number of anonymized tuples. A penalty of 1 is assigned to each tuple whose class label differs from the majority class label of its equivalence class. If the tuple's class label matches the majority, no penalty is given. The CM is related to but distinct from the concept of pureness. Unlike pureness, which considers all possible attribute value combinations, the CM specifically evaluates the class label uniformity within each equivalence class. Pureness is considered more suitable for evaluating how often an anonymous counterfactual explanation provides correct advice because it takes into account the entire range of possible attribute combinations, rather than just the observed instances~\cite{goethals2023privacy}.

\sstitle{RDT-Fidelity}
Olatunjii et al.~\cite{OlatunjiRFK23} describes a metric for measuring the quality of explanations for model predictions through a metric called faithfulness. Faithfulness indicates how well an explanation approximates the model's behavior. Since a ground truth for explanations is often unavailable, the measure used is RDT-Fidelity (grounded in rate-distortion theory~\cite{funke2022z}), which assesses faithfulness by comparing the model's original and new predictions. The fidelity score is calculated as follows:
\[ \mathcal{F}(\mathcal{E}_X) = \mathbb{E}_{Y_{\mathcal{E}_X} | Z \sim \mathcal{N}} \left[ 1_{f(X)=f(Y_{\mathcal{E}_X})} \right] \]
Here, \( \mathcal{E}_X \) represents the explanation, \( f \) is the model function (like a Graph Neural Network), \( X \) is the original input, \( \mathcal{M}(\mathcal{E}_X) \) is the explanation mask applied to \( X \), \( Z \) is noise drawn from distribution \( \mathcal{N} \), and \( \tilde{I}_{\mathcal{E}_X} \) is the perturbed input defined by:
\[ \tilde{I}_{\mathcal{E}_X} = X \odot \mathcal{M}(\mathcal{E}_X) + Z \odot (1 - \mathcal{M}(\mathcal{E}_X)), Z \sim \mathcal{N}, \]
where \( \odot \) denotes element-wise multiplication and \( 1 \) represents a matrix of ones of appropriate size. The score reflects how often the model's predictions are unchanged despite perturbations to the input, which would suggest that the explanation is effectively capturing the reasoning behind the model's predictions.

\sstitle{Sparsity}
Olatunjii et al.~\cite{OlatunjiRFK23} argues that a complete and faithful explanation to the model should inherently be sparse, focusing only on a select subset of features that are most predictive of the model's decision. The measurement of sparsity is done using an entropy-based definition which can be applied to both soft and hard explanation masks. The sparsity of an explanation is quantified by the entropy \( H(p) \) over the normalised distribution \( p \) of the explanation masks, calculated using the formula~\cite{funke2022z}:
\[ H(p) = -\sum_{f \in M} p(f) \log p(f) \]
Here, \( M \) represents the set of features and \( \log(|M|) \) bounds the entropy. A lower entropy value implies a sparser explanation.

\subsubsection{Information loss}
Excessive anonymisation often results in the loss of valuable information. As the level of anonymisation increases, the data utility typically decreases, hindering certain types of analysis or yielding outcomes that are biased or inaccurate.

\sstitle{Normalised Certainty Penalty (NCP)}
It quantifies the information loss that occurs when attributes are anonymised~\cite{goethals2023privacy}. NCP is higher for attributes that, when generalised, encompass a wide range of possible values, indicating greater information loss:
For numerical quasi-identifiers in an equivalence class \( G \), NCP is calculated using:
$ \text{NCP}_{A_{num}}(G) = \frac{max^G_{A_{num}} - min^G_{A_{num}}}{max^{A_{num}} - min^{A_{num}}} $.
For categorical quasi-identifiers, $\text{NCP}_{A_{cat}}(G)$ is $0$ if $|A^G| = 1$ and $\frac{|A^G|}{|A|}$ otherwise.
The overall NCP for an equivalence class \( G \) across all quasi-identifier attributes is the weighted sum:
\[ \text{NCP}(G) = \sum_{i=1}^{d} w_i \cdot \text{NCP}_{A_i}(G) \]
where \( d \) is the number of quasi-identifiers, \( A_i \) is the \( i^{th} \) attribute with weight \( w_i \), and \( \sum w_i = 1 \).
Higher NCP values indicate a greater degree of generalization and more information loss. This metric helps in assessing the balance between data privacy and utility.

\sstitle{Discernibility}
The discernibility metric \( C_{DM}(g, k) \), which is used to measure the penalties on tuples in a dataset after k-anonymization, reflecting how indistinguishable they are post-anonymization~\cite{goethals2023privacy}. The goal is to maintain discernibility between tuples within the constraints of a given privacy level k. The metric is defined as:
\[ C_{DM}(g, k) = \sum_{VE \,s.t.\, |E| \geq k} |E|^2 + \sum_{VE \,s.t.\, |E| < k} |D||E| \]
Here, \( E \) denotes the equivalence class of the tuple, and \( D \) represents the entire dataset. A successfully anonymized tuple (with an equivalence class larger than k) incurs a penalty equivalent to the square of the equivalence class size, while a suppressed tuple (with an equivalence class smaller than k) incurs a penalty proportional to the size of the dataset multiplied by the equivalence class size. The metric has been critiqued for not considering how closely the anonymized instances resemble the original data~\cite{goethals2023privacy}. The Normalized Certainty Penalty (NCP) is suggested as a more appropriate metric for gauging the actual information loss in the process of anonymizing counterfactual explanations.

\sstitle{Error in private approximation}
Patel et al.~\cite{patel2022model} proposes a metric to measure the error caused by randomness added when privately minimizing \(\mathcal{L}(\cdot)\) for protecting \(X\) as the expected deviation of the randomized explanation from the best local approximation. More formally, the approximation loss is defined as:
\[
\mathcal{E}(\hat{\phi}, \mathcal{Z}, f(X)) \triangleq \mathbb{E} [\mathcal{L}(\hat{\phi}, \mathcal{Z}, f(X)) - \mathcal{L}(\phi^*, \mathcal{Z}, f(X))].
\]

\sstitle{Explanation Intersection}
Olatunjii et al.~\cite{OlatunjiRFK23} measures the percentage of bits in the original explanation that is retained in the privatised explanation after using differential privacy~\cite{funke2022z}.

\subsubsection{Privacy degree}
Degree of privacy refers to the level of privacy protection, which can be measured in different aspects.

\sstitle{k-anonymity degree}
$k$-anonymity refers to the number of individuals in the training dataset to whom a given explanation could potentially be linked~\cite{goethals2023privacy}. This concept is grounded in the principle of k-anonymity, which ensures that a person's information is indistinguishable from at least k-1 other individuals.

\sstitle{Information leakage}
For a sequence of queries \( \mathbf{z_1}, \mathbf{z_2}, \ldots, \mathbf{z_k} \), the algorithm is (\(\hat{\varepsilon}, \hat{\delta}\))-differentially private if the probability ratio of generating an explanation for any of the queries is bounded by \( e^{\hat{\varepsilon}} \) times the probability of the explanation under a differentially private model \( f \), plus a term \( \hat{\delta} \)~\cite{patel2022model}:
\[ Pr_{i=1..k}\hat{\phi}(\mathbf{z_i}, X, f_D(X)) \leq e^{\hat{\varepsilon}} \cdot Pr[\hat{\phi}(\mathbf{z_i}, X, f'_D(X)) : \forall i] + \hat{\delta}, \]
where \( \hat{\varepsilon} \leq \varepsilon \) and \( \hat{\delta} \leq \delta \), and at least one of the inequalities is strict.
Intuitively, this means that even if an adversary has access to the model explanations, they would not gain any additional information that could help in inferring something about the training data beyond what could be learned from the model predictions alone.

\sstitle{Privacy budget}
Patel et al.~\cite{patel2022model} measures the allocation of a privacy budget for an explanation dataset that comprises a sequence of queries. The total privacy budget for all queries is fixed at (\( \varepsilon, \delta \)), with a stricter privacy requirement (\( \varepsilon_{min}, \delta_{min} \)) set for each individual query to prevent significant information leakage. The explanation algorithm must ensure global privacy adherence by not exceeding the overall privacy budget across all queries. This means that the probability of the algorithm providing explanations within certain sets \( S_1, S_2, \ldots, S_k \) should be less than or equal to the product of \( e^\varepsilon \) and the probability of these explanations under a differentially private algorithm, plus \( \delta \). Furthermore, for every individual query \( \mathbf{z_j} \), the probability should be within \( e^{\varepsilon_{min}} \) times the differentially private algorithm probability plus \( \delta_{min} \). The goal is to create an explanation algorithm that can address as many queries as possible without exceeding the designated privacy budget and while still providing quality assurances.

\subsubsection{Attack success}

Measuring the success of privacy attacks is a cornerstone to evaluate the effectiveness of designed attacks, which in turn reflect the risk of a given XAI system.

\sstitle{Precision/Recall/F1}
In terms of attribute reference attacks~\cite{duddu2022inferring}, \emph{Precision} is the percentage of the positive attributes inferred by an attack being indeed positive according to the ground truth. \emph{Recall} is the percentage of relevant instances of positive attributes being identified by an attack. Lastly, the \emph{F1 Score} is the harmonic mean of precision and recall, calculated as \(2 \times \frac{\text{precision} \times \text{recall}}{\text{precision} + \text{recall}}\), which balances precision and recall; it reaches its best value at 1 (perfect precision and recall) and worst at 0, when either precision or recall is zero.

\sstitle{Balanced accuracy (BA)}
This metric measures the accuracy of attack (e.g. membership inference), on a balanced dataset of members and non-members~\cite{pawelczyk2023privacy,liu2024please}: 
\[
BA = \frac{TPR + TNR}{2}
\]
where TPR is true-positive rate (true membership prediction) and TNR is true-negative rate (true non-membership prediction).

\sstitle{ROC/AUC}
ROC (Receiver Operating Characteristic) curve and AUC (Area Under the Curve) are metrics adapted from machine learning to measure the success of privacy attacks, such as re-identification or membership inference attacks~\cite{pawelczyk2023privacy}. The ROC curve plots the TPR against the FPR at various threshold settings, providing a visual representation of an attack's ability to distinguish between different classes (e.g., members vs. non-members in a dataset). The AUC, a single value derived from ROC, quantifies the overall effectiveness of the attack across all thresholds~\cite{huang2023accurate}.

\sstitle{TPR at Low FPR}
TPR at Low FPR~\cite{liu2024please,huang2023accurate} is used to measure attack performance at a fixed FPR (e.g., 0.1\%).
Evaluating the True Positive Rate (TPR) at low False Positive Rates (FPR) is essential in scenarios where the cost of false positives is high, because it ensures that the positive results are both accurate and reliable. Low FPR evaluation is crucial particularly in imbalanced datasets, where false positives can outnumber true positives. For example, if a membership inference attack can pinpoint even a minuscule fraction of the training dataset with high precision, then the attack ought to be deemed effective~\cite{pawelczyk2023privacy}.

\sstitle{Mean Absolute Error (MAE)}
Denoted as \( \ell_1 \) loss, it quantifies the average magnitude of the errors between the reconstructed inputs \( \hat{x} \) and the original inputs \( x \):
\[ \ell_1 (\hat{x}, x) = \frac{1}{mn} \sum_{j=1}^{m} \sum_{i=1}^{n} | \hat{x}_i^j - x_i^j |, \]
where \( m \) is the number of samples in the validation dataset \( X_{\text{val}} \) and \( n \) is the number of features in the dataset~\cite{luo2022feature}. 

\sstitle{Success Rate (SR)}
The Success Rate (SR) is defined as the ratio of the count of successfully reconstructed features to the total number of features across all samples:
\[ SR = \frac{|\hat{X}_{val} \neq \perp|}{mn}, \]
where \( |\hat{X}_{val} \neq \perp| \) denotes the number of features that are not equal to a specific value \( \perp \) (represents a reconstruction failure or a null value), \( m \) is the number of samples, and \( n \) is the number of features. This metric quantifies the portion of the dataset \( X_{val} \) where features are correctly reconstructed by the attack.

\sstitle{Model agreement}
In the context of model extraction attacks, Wang et al.~\cite{wang2022dualcf} uses \emph{agreement} as a measure for comparing the behavior of a high-fidelity model \( h_\phi \) to a target model \( f_\theta \). The agreement is defined as the average number of predictions where \( f_\theta \) and \( h_\phi \) coincide, over an evaluation set of size \( n \):
\[ \text{Agreement} = \frac{1}{n} \sum_{i=1}^{n} 1_{f_\theta(x_i) = h_\phi(x_i)}. \]
A higher agreement indicates that the substitute model \( h_\phi \) is more similar to the original model \( f_\theta \). When comparing two model extraction methods with the same agreement, the one with the lower standard deviation is preferred.

\sstitle{Average uncertainty reduction}
Ferry et al.~\cite{ferry2023probabilistic} evaluates the effectiveness of a data reconstruction attack. Consider a deterministic dataset \( \mathcal{D}^{Orig} \) composed of \( n \) samples each with \( d \) features, which is used to train a machine learning model \( M \). Let \( \mathcal{D}^M \) represent a probabilistic dataset that is reconstructed from \( M \). By its design, \( \mathcal{D}^M \) should align with \( \mathcal{D}^{Orig} \). The degree to which the reconstruction is accurate is measured by the reduction in uncertainty across all features of all samples in the dataset, on average:
\[ Dist(\mathcal{D}^M, \mathcal{D}^{Orig}) = \frac{1}{n \cdot d} \sum_{i=1}^{n} \sum_{k=1}^{d} \frac{H(\mathcal{D}^M_{i,k})}{H(\mathcal{D}_{i,k})} \]
Here, the random variable \( \mathcal{D}_{i,k} \) symbolizes an uninformed reconstruction, evenly distributed across all conceivable values of feature \( k \) of attribute \( a_k \), and \( H \) denotes the Shannon entropy.
Lower values of \( Dist(\mathcal{D}^M, \mathcal{D}^{Orig}) \) reflect superior reconstruction attacks.

\section{Future Research Directions}
\label{sec:future}

\subsection{Ethical Implications}

The push for explainable AI has led to the development of tools and startups like MS InterpretML, Fiddler Explainable AI Engine, IBM Explainability 360, Facebook Captum AI, and H2O Driverless AI~\cite{gade2019explainable}. Our survey explores the privacy risks of making ML models explainable, highlighting the potential for malicious exploitation of these explanations, especially for high-risk data such as medical records and financial transactions. This raises concerns about the conflict between the right to explain ML models~\cite{goodman2017european} and user privacy, necessitating discussions involving legal experts and policymakers~\cite{banisar2011right}. Additionally, the tension between explainability and privacy may disproportionately impact minority groups by either exposing their data or providing lower-quality explanations~\cite{shokri2021privacy}.

This survey contributes to a broader research agenda on AI transparency and privacy, sparking discussions among scholars focused on AI governance. Although the trade-off between privacy and explainability is not a novel issue in legal discussions~\cite{kaur2020interpreting}; we remain hopeful about developing explanation methodologies that safeguard user privacy, albeit potentially at the expense of explanation quality. 
While explanation quality is subjective, one thing is clear: explanations that fail to reveal useful model insights while protecting user data are likely less beneficial to end-users~\cite{shokri2021privacy}.

\edit{Looking into the future, the ethical implications of privacy-preserving techniques include balancing privacy protection with transparency and fairness~\cite{hu2022protecting}. Techniques like differential privacy and federated learning secure data by adding noise or decentralising processing, but they can reduce model accuracy and transparency, complicating trust and understanding~\cite{liu2024matrix,liu2024guaranteeing}. These methods can also introduce biases, affecting certain groups disproportionately and amplifying discrimination~\cite{mi2024towards}. Ensuring informed consent and user autonomy is crucial, necessitating clear communication about how these techniques impact data use and model performance~\cite{zhang2024does}.
}

\subsection{Regulatory Compliance}

\edit{Privacy attacks on model explanations pose significant challenges under regulatory frameworks like the GDPR, which emphasise the protection of personal data and transparency in automated decision-making. Such attacks can lead to unauthorised data disclosure, complicating compliance with GDPR's requirements for data subject rights, including access and erasure~\cite{nguyen2022survey,trung2024fastfedul}. Additionally, privacy-preserving techniques that obscure model explanations may hinder transparency, making it difficult for organisations to demonstrate compliance and for individuals to understand AI decisions, thereby affecting accountability~\cite{liu2024dynamic}. Moreover, these techniques must balance privacy and utility, as overly restrictive measures can impact the effectiveness and fairness of AI systems, posing further challenges for legal and ethical standards~\cite{zhang2024does}.
}

\subsection{Privacy Tradeoffs}

Li et al.~\cite{li2023balancing} discusses the impact of differential privacy on the interpretability of deep neural networks.
It examines how injected noise into the model parameters affects the gradient-based interpretability method. The analysis reveals that while noise in the fully connected layer directly affects the feature map used for interpretability, noise in the convolutional layer alters the output of the activation function, thus impacting the feature map indirectly.
Chang et al.~\cite{chang2021privacy} examines the relationship between algorithmic fairness and privacy. It points out that while fair machine learning models strive to reduce discrimination by equalising behaviour across different groups, this process can alter the influence of training data points on the model, leading to uneven changes in information leakage. Fair algorithms may inadvertently memorise and leak more information about under-represented subgroups in an attempt to equalise errors across different groups based on protected attributes.
The findings indicate a trade-off where achieving fairness for protected or unprivileged groups amplifies their privacy risks. Moreover, the greater the initial bias in the training data, the higher the privacy cost when making the model fair for these groups. These findings are relevant to model explanations, which also impact fairness~\cite{dodge2019explaining,zhang2018fairness}.

\subsection{Underexplored Privacy Attacks}

Aivodji et al.~\cite{aivodji2022fooling} present techniques for manipulating and detecting manipulation of SHAP values. To manipulate SHAP values, a brute-force sub-sampling method is used to minimise the differences in SHAP values, with a clever re-weighting strategy to make the sampling appear legitimate. 
Detection of such manipulation employs statistical tests to compare model outputs from manipulated and unmanipulated samples~\cite{frye2020shapley}. 
Slack et al.~\cite{slack2020fooling} outlines a framework for constructing adversarial classifiers that deceive post hoc explanation techniques, such as LIME and SHAP. The framework produces an adversarial classifier that mimics the biased classifier on real distribution data but reverts to unbiased predictions on out-of-distribution (OOD) data~\cite{mittelstadt2019explaining}. 
Regarding data reconstruction attacks, an interesting direction is to utilize the inner workings of learning algorithms in some interpretable models (e.g. decision tree) to reduce the entropy of probabilistically reconstructed datasets. For example, since greedy algorithms for constructing decision trees select features based on Gini impurity, we can identify and discard certain attribute combinations that do not contribute to an optimal decision tree~\cite{ferry2023probabilistic}.

\subsection{Underexplored Model Explanations}

Gillenwater et al.~\cite{gillenwater2021differentially} introduces a novel method for computing multiple quantiles in sensitive data with differential privacy. Traditional methods compromise on accuracy by either splitting the privacy budget across quantiles or inefficiently summarizing the entire distribution. The proposed approach uses an exponential mechanism to estimate multiple quantiles efficiently, achieving better accuracy and efficiency compared to existing methods. 
This is particularly relevant because there are emerging explainability measures based on quantiles~\cite{ghosh2022faircanary,li2023explainable,merz2022interpreting}.
Alvarez et al.~\cite{alvarez2018towards} proposes the concept of self-explaining models that incorporate interpretability from the onset of learning. The authors design self-explaining models in a stepwise manner, starting from simple linear classifiers and advancing to more complex structures with built-in interpretability~\cite{zhang2022protgnn}. They introduce specialized regularization techniques to maintain faithfulness and stability.
Olatunji et al.~\cite{OlatunjiRFK23} pioneer the examination of privacy risks tied to feature explanations in graph neural networks (GNNs), presenting scenarios where adversaries attempt to unveil hidden relationships within the data, despite having limited access to the network's structure~\cite{khosla2022privacy}. The paper delves into various explanation methods for GNNs such as gradient-based, perturbation-based, and surrogate methods. Furthermore, it outlines potential adversarial attacks aimed at exploiting these explanations to compromise privacy and introduces a novel defense mechanism based on perturbing explanation bits to adhere to differential privacy standards.
Other works~\cite{tiddi2022knowledge,rajabi2022knowledge} examine the role of knowledge graphs as model explanations, positing that integrating structured, domain-specific knowledge can lead to more understandable, insightful, and trustworthy AI systems. However, knowledge graphs can be used to fuel privacy attacks such as de-anonymisation and membership inference~\cite{qian2017social,wang2021membership}.

\subsection{Underexplored Data Modalities}

\sstitle{Graph Data}
The rapid development in the area of graph neural networks (GNNs)~\cite{huynh2021network, duong2022deep,nguyen2014reconciling,nguyen2015smart,hung2019handling} highlights a special treatment for GNN explainability~\cite{wu2020comprehensive}.
Yuan et al.~\cite{yuan2022explainability} discuss explainability methods specifically designed for Graph Neural Networks (GNNs) such as gradients/features-based, perturbation-based, surrogate, and decomposition methods. Prado et al.~\cite{prado2023survey} provides a comprehensive overview of graph counterfactual explanations for GNNs. Privacy attacks on GNNs are also an emerging direction~\cite{dai2022comprehensive}.

\sstitle{Audio Data}
Audio signals consists of speech signals and other non-speech audio signals. Speech processing involves tasks like automatic speech recognition, speaker identification, and paralinguistic information recognition, while non-speech audio signal processing contains many more applications, such as human heart sound analysis, bird sound analysis, and environmental sound classification. Current research have separately focused on data / model privacy and explanation approaches~\cite{ren2023anoverview,li2021robust,carlini2018audio,abdullah2021sok}. While explainable models are essential for audio-based healthcare applications~\cite{ren2022heartsound,ren2020generating,chang2022example}, there is still a large gap to further explore the privacy risks of audio-based model explanations.

\subsection{Privacy-Preserving Models}

Exploring how privacy-preserving models, such as differentially private decision trees, reduce the success of privacy attacks represents a valuable research direction~\cite{ferry2023probabilistic}. Li et al.~\cite{li2023balancing} presents an Adaptive Differential Privacy (ADP) mechanism aimed at improving the interpretability of machine learning models without compromising privacy. This mechanism selectively injects noise into the less critical weights of a model's parameters, thereby preserving the interpretability of important features which conventional differential privacy methods may obscure.

\subsection{Privacy-Protecting Explanations}

Using model explanations to counter adversarial attacks is a novel direction. Belhadj et al.~\cite{belhadj2021fox} outlines a framework (called FOX) to safeguard social media users' privacy by using adversarial reactions to trick classifiers. It constructs a dataset of social media interactions, employs an explainability tool to extract influential adversarial features, and filters them to create a robust list. These features are then used to generate adversarial reactions, aiming to mislead the classifier away from the correct classification and towards a predetermined label, thus preserving the user's privacy.

\subsection{Time Complexity}

Time complexity is crucial in privacy attacks on model explanations. Fast run-time methods pose higher risks by enabling rapid exploitation, while more complex iterative attacks are less practical due to longer execution times. The feasibility of these attacks depends on computational resources and scalability. Effective countermeasures must balance protection and performance to mitigate risks from fast, real-time attacks. Unfortunately, only a few works thoroughly discuss time complexity such as Shapley approximation~\cite{jia2019towards} and DP-quantiles~\cite{gillenwater2021differentially}.

\section{Conclusions}
\label{sec:conclusion}

\sstitle{Summary}
As the prevalence of model explanations grows, there is an emerging interest in understanding its repercussions, including aspects of fidelity, fairness, stability, and privacy. This survey offers a thorough investigation into the latest privacy-centric attacks on model explanations, establishing a comprehensive classification of these attacks based on their traits. Furthermore, it delves deeply into the present advanced research on defensive strategies and privacy-focused model explanations, uncovering common privacy design approaches and their variations.

Our survey also highlights several unresolved issues that demand additional inquiry. Primarily, it points out the current research's limited scope, which predominantly focuses on membership inference attacks, counterfactual explanations, and differential privacy. It suggests that numerous widely-used algorithms and models, in terms of their real-world implementation and relevance, deserve more detailed scrutiny. Secondly, there's a noticeable lack of deep theoretical insight into the origins of privacy breaches, impacting both the development of protective measures and the comprehension of privacy attack limitations. Although experimental research into the determinants of privacy breaches has yielded valuable knowledge, there's a scarcity of studies evaluating attacks under realistic conditions, considering dataset size and actual deployment. As the field continues to explore the privacy implications of model explanations, this survey aims to serve as a crucial resource for interested readers eager to contribute to this trend.

\sstitle{Challenges}
The challenges for new work in this field, as highlighted in the survey, include:
\begin{compactitem}
\item \emph{Balancing Transparency and Privacy:} Providing detailed explanations improves transparency but increases the risk of privacy breaches by revealing sensitive information embedded in the training data.
\item \emph{Granularity of Explanations:} Detailed explanations can lead to direct inferences about data points, making it challenging to protect privacy without losing interpretability.
\item \emph{Understanding Privacy Leaks:} Identifying the causes of privacy leaks through model explanations is complex and requires thorough investigation of different explanation methods and their vulnerabilities.
\item \emph{Diverse Attack Models:} Developing comprehensive defenses against a wide range of privacy attacks, including membership inference, model inversion, and reconstruction attacks, is necessary but challenging due to the evolving nature of these attacks.
\item \emph{Countermeasure Effectiveness:} Evaluating and improving the effectiveness of countermeasures, such as differential privacy and perturbation techniques, to ensure they do not compromise the utility of model explanations.
\item \emph{Dynamic Interaction Scenarios:} Assessing the impact of repeated interactions between adversaries and the model in dynamic settings adds complexity to designing robust privacy-preserving methods.
\item \emph{Interpretable Surrogates:} Surrogate models used for providing explanations can themselves become targets for privacy attacks, necessitating additional safeguards.
\item \emph{Scalability and Practicality:} Implementing privacy-preserving techniques in real-world must balance scalability and practicality without significantly affecting model performance.
\end{compactitem}



\end{document}